\documentclass[12pt]{article}
\usepackage{a4,amsmath,epsfig,amssymb}
\usepackage[latin1]{inputenc}
\usepackage{graphicx}
\usepackage{caption}
\usepackage{subcaption}
\usepackage{feynmp}
\begin{document}
\bigskip\bigskip

\begin{center}
{\large \bf Predictions and Tests \\
for Semi-leptonic Decays of Beauty Hadrons} 
\end{center}
\vspace{8pt}
\begin{center}
\begin{large}

E.~Di Salvo$^{a,b,c,}$\footnote{elvio.disalvo@ge.infn.it},
Z.~J.~Ajaltouni$^{a,}$\footnote{ziad@clermont.in2p3.fr}
and
F.~Fontanelli$^{b,c,}$\footnote{In memory: he gave important contributions to the article} 
\end{large} 

\bigskip
$^a$ 
Laboratoire de Physique de Clermont - UCA \\
4 Av. Blaise Pascal, TSA60026, F-63178 Aubi\`ere Cedex, France\\

\noindent  
$^b$ 
Dipartimento di Fisica 
 Universit\'a di Genova \\
Via Dodecaneso, 33, 16146 Genova, Italy
 
\noindent  
$^c$ 
I.N.F.N. - Sez. Genova,\\
Via Dodecaneso, 33, 16146 Genova, Italy \\  

\noindent  

\vskip 1 true cm
\end{center} 
\vspace{1.0cm}

\vspace{6pt}
\begin{center}{\large \bf Abstract}

We make predictions and propose tests for the semi-leptonic decays $B\to D^{(*)} l \bar{\nu}_l$ and $\Lambda_b\to \Lambda_c l \bar{\nu}_l$, with $l$ = $e$, $\mu$ and $\tau$. Preliminarily, we study such processes according to the $\ell-s$ scheme, which helps interpreting the results of previous analyses. Then, we write the angular distributions of those decays in the helicity formalism and suggest for each decay some tests for distinguishing among the various Dirac operators which may explain the tensions with the standard model. In particular, we propose, for $l$ = $e$ and $\mu$, to study the fully differential distribution. In the cases of the $B\to D^*$ and $\Lambda_b$ decays, important physical information can be extracted even after integrating over the momentum of the charged lepton, which is quite suitable for $l$ = $\tau$. Especially, the decay angular distribution of polarized $\Lambda_b$ depends on five observables, which assume different values according to the various new-physics interactions; one of these observables is related to the longitudinal polarization of the $\tau-\bar{\nu}_\tau$ system, which can be compared with the $D^*$ longitudinal polarization in the $B\to D^*$ decay. Moreover, we analyze the $\tau$ anomaly in the branching ratios of the semi-leptonic decays; to this end, we make some assumptions, which have a remarkable predictive power and lead to results in agreement with previous theoretical calculations. Thanks to these assumptions, we combine our present analysis of the $\Lambda_b$ decay with the previous ones of the $B\to D^{(*)}$ decays, getting strong constraints on the parameters which characterize the different new-physics operators and an indication on the most likely one. 

\end{center}

\vspace{10pt}

\centerline{PACS numbers: 13.30.Ce, 12.15.-y, 12.60.-i}

\newpage

\section{Introduction}  

Tensions with the standard model (SM) are exhibited in Babar, LHC and Belle measurements of the semi-leptonic decays 
\begin{equation}
B\to D^{(*)} \ell \bar{\nu}_{\ell} (\tau \bar{\nu}_{\tau}), \label{bdt}
\end{equation}
with $\ell$ = $e$ or $\mu$[1-9]. Indeed, comparing the $\mu$- to the $\tau$- decay contradicts the lepton flavor universality (LFU), which is a consequence of the SM. More precisely, the observables, which are in contrast with the SM predictions, are the ratios 
\begin{equation}
{\cal R}_{D^{(*)}}= \frac{{\cal B} (B\to D^{(*)} \tau \bar{\nu}_{\tau})}
{{\cal B}(B\to D^{(*)} \mu \bar{\nu}_{\mu})}. \label{rtB} 
\end{equation}
The result - which is confirmed by a recent measurement\cite{lhc4} - appears promising in the search for physics beyond the standard model, as witnessed by the numerous theoretical efforts that have been performed in the last few years. In this connection, we cite some recent theoretical articles[10-30] and reviews[31-36].
 
This suggests that an interaction beyond the SM occurs at short distances, at most $O(10^{-3} fm)$, since the final state (long-ranged) interactions between leptons and hadrons are negligibly small. Therefore the attention is focused on the basic process
\begin{equation}
b\to c \ell \bar{\nu}_{\ell} (\tau \bar{\nu}_{\tau}), \label{ccu}
\end{equation}
which is assumed to be responsible for the anomaly. People looked for confirmations to this conjecture in other decays, in particular\cite{dub},  
\begin{equation}
B_c\to J/\psi (\eta_c) \ell \bar{\nu}_{\ell} (\tau \bar{\nu}_{\tau}).  \label{bpsi}
\end{equation}
Indeed, the measurement\cite{lh18} of ${\cal R}_{J/\psi}$ - which is defined analogously to (\ref{rtB}) - supports the assumption, although with large uncertainties, which stimulated new suggestions in this sense[21,27,39-50].
Further indications in favor of a new physics (NP) may come from the longitudinal components of the polarizations of $\tau$ ($P^\tau_{D^*}$\cite{bel1,bel2}) and of $D^*$ ($F^L_{D^*}$\cite{bel3,kkr}) in the decay (\ref{bdt}); a number of articles are dedicated to the topics[14,17,21,24,42,54-59]. 

As regards the decay
\begin{equation}
\Lambda_b\to \Lambda_c \ell \bar{\nu}_{\ell} (\tau \bar{\nu}_{\tau}), \label{lmb}
\end{equation}
it attracted the interest of several authors[10,11,23,60-69].
The experimental value of ${\cal R}_{\Lambda_c}$\cite{cern} seems to contradict the trend shown by the results of ${\cal R}_{D^{(*)}}$, as confirmed by sum rules\cite{fb,ymn,igw}; this measurement is not conclusive, owing to the statistical and systematic errors and to the uncertainties on the elaboration of data\cite{BLN}. 

Also the ratio ${\cal R}_{X_c}$, relative to the inclusive decay
\begin{equation}
B\to X_c \ell \bar{\nu}_{\ell} (\tau \bar{\nu}_{\tau}), \label{incl}
\end{equation}
was investigated years ago\cite{bAl}, without providing a definite answer. 

In order to solve definitively the puzzle, the measurement of different observables is proposed. Indeed, till now, the hints to NP come essentially from lifetime ratios of the type ${\cal R}$, analogous to (\ref{rtB}). It appears suitable to look for confirmations of this anomaly from single semi-leptonic decays, and not only with $\tau$ leptons in the final state, but also with muons, which are more easily detectable. In particular, measurements of angular and $q^2$-distributions for the decays above are proposed[11,14,16,18,22,23,29,33,41,54-56,63,71-74] and the muon-electron universality is tested\cite{prm,whe}. Moreover, since the SM excludes for such decays CP violation (CPV) and time reversal violation(TRV), tests for these effects are suggested\cite{hgw,gjl}. 

However, very few papers are dedicated to a simultaneous analysis of the meson and baryon semi-leptonic decays\cite{ymn,gro,hu1,wei,celi,ikw}. In the present article, we show an intriguing interplay among such decays.
In particular, we make some assumptions, whose consequences agree with previous theoretical calculations and give rise to predictions about the ratios ${\cal R}$ for different semi-leptonic decays. Thanks to these assumptions, we compare our results on the $\Lambda_b$ decay with those by other authors on the $B$ decays, getting stringent constraints about the kind of interaction that cause the tensions of data with the SM.

We are mainly concerned with the decays (\ref{lmb}), which present some advantages. It is well-known that the determination of the momentum distribution of the $\tau$ lepton presents some difficulties\cite{gjl,btt2,aln}; but, as we shall see, thanks to the secondary decay of $\Lambda_c$ (typically to $\Lambda \pi$), important physical information can be drawn from the angular distribution, even after integrating over the $\tau$  momentum. Also, we single out five coefficients of such a distribution, which assume different values according to the Dirac operator which describes NP; one of them is related to the longitudinal polarization of the $D^*$ found in some of the decays (\ref{bdt}), which implies further constraints on NP.

Besides, the fully differential angular distributions are useful in searching for hints at NP in semi-leptonic decays where a light lepton is involved in the final state. Therefore, we study them in detail, both for meson and baryon decays. To this end, we observe that some helicity amplitudes, which describe the decays of the intermediate bosons to the $l-\bar{\nu_l}$ system, are common to the $B$ and $\Lambda_b$ decay and, as we shall illustrate, they give rise to some predictions as regards the $B$ decays. Furthermore, such distributions allow to elaborate tests for TRV, but, if one integrates over the lepton momentum, only the $\Lambda_b$ decay is still sensitive to that violation.

We concentrate mainly on the decays (\ref{bdt}) and (\ref{lmb}), for which both experimental data  and precise evaluations of the SM predictions are presently  available; however, some considerations and conclusions can be extended also to other decays, among which (\ref{bpsi}) and (\ref{incl}). We adopt mainly the helicity representation[80-84], but also the $\ell-s$ scheme\cite{kosc} is suitable for interpreting in a simple way some features of previous analyses of $B$ decays.

The article is organized as follows. In Sect. 2, we expose the general formalism for the semi-leptonic decays. In Sect. 3, we illustrate some features about the decays (\ref{bdt}) and (\ref{lmb}) in the $\ell-s$ scheme. Sects. 4 and 5 are dedicated to the study of the angular distributions of those decays, in the helicity formalism. In Sect. 6, we calculate the helicity amplitudes for the decay (\ref{lmb}), which are in part applicable also to the decays (\ref{bdt}). Incidentally, we stress that the results of Sects. 4 to 6 may be applied both to $\tau$ and light leptons in the final state. Sect. 7 is devoted to a phenomenological analysis, where, starting from some assumptions, we deduce important predictions; moreover, we explain some results of previous analyses. Last, Sect. 8 is dedicated to a summary and to a discussion of our results.

\section{General Formalism}

We develop, here and in Sects. 3 to 6, a formalism for the differential decay widths of the decays (\ref{bdt}) and (\ref{lmb}). Similar expressions are found in the literature, both for the $B$[16,18,21,22,25,27,29,54-56,58,74,86-90]  and for the $\Lambda_b$\cite{phn,gjl,dt2,bkt,gu8,huea} semi-leptonic decays; however, we unify, as far as possible, the descriptions of the two kinds of decays and propose alternative tests for getting possible signals of NP. 

\subsection{Differential Decay Widths}

Consider a decay of the type 
\begin{equation}
R_1 \to R_2 l \bar{\nu}_l,
\end{equation}
with $R_1$ = $B$ or $\Lambda_b$ and $R_2$ = $D$, $D^*$ or $\Lambda_c$. The differential decay width reads as
\begin{equation}
d\Gamma=\frac{1}{2M_1} T d\Phi_3, ~~~~ T = \sum_{M,M'}\rho_{MM'}\sum_f {\cal M}_M^{fi}{\cal M}_{M'}^{fi*}. \label{dga}
\end{equation} 
Here $M_1$ denotes the mass of the parent hadron $R_1$, $\rho_{M,M'}$ its spin density matrix and ${\cal M}_M^{fi}$ the matrix element of the decay; moreover, 
\begin{equation}
d\Phi_3 = \frac{1}{2^3} \frac{1}{(2\pi)^5}\frac{p_2}{M_1}d\Omega_2 p_l dQ d\Omega_l;                 
\end{equation}
$Q$ = $(q^2)^{1/2}$ is the effective mass of the $l-{\bar \nu}_l$ system and ($p_2$, $\Omega_2$) and ($p_l$, $\Omega_l$) denote the momenta and directions, respectively, of $R_2$ in the $R_1$ ({\it canonical}) rest frame (RF) and of $l$ in the  $l-{\bar \nu}_l$  ({\it helicity}) RF. In particular, we have adopted the notation
\begin{equation}
\Omega_2 \equiv (\theta_2,\phi_2),
\end{equation}
where $\theta_2$ and $\phi_2$ are, respectively, the polar and azimuthal angle of the momentum of $R_2$, an analogous notation holding for $\Omega_l$. In this connection, it is worth recalling the definition of the helicity frame\cite{jw,ch0}: it is characterized by the mutually orthogonal unit vectors
\begin{equation}
{\hat z} = -\frac{\vec{p}_2}{p_2}, ~~~ {\hat y} = \frac{{\hat Z}\times{\hat z}}{|{\hat Z}\times{\hat z}|}, ~~~
{\hat x} = {\hat y}\times{\hat z},  \label{hrf}
\end{equation}
where $\vec{p}_2$ is the momentum vector of $R_2$ in the canonical frame and ${\hat Z}$ is the unit vector along the $z$-axis of the canonical frame.

If $R_2$ is unstable, and a decay of the type $R_2\to R_3 \pi$ is observed ($D^* \to D \pi$ or $\Lambda_c \to \Lambda\pi$), the phase space $d\Phi_3$ has to be replaced in Eq. (\ref{dga}) by
\begin{equation}
d\Phi_4 = 
d\Phi_3 \frac{1}{2\pi} \frac{1}{(4\pi)^2}\frac{p_3}{M_2}d\Omega_3.
\end{equation}
Here $p_3$ and $\Omega_3$ denote, respectively, the modulus and the direction of the momentum of $R_3$ in the $R_2$ ({\it helicity}) RF, whose unit vectors are defined by  Eqs. (\ref{hrf}), except for 
\begin{equation}
{\hat z} = \frac{\vec{p}_2}{p_2}. \label{hrf1}
\end{equation}
For later convenience, we express the unit vectors of the helicity RF of  $R_2$ in terms of $\theta_2$ and $\phi_2$:
\begin{eqnarray} 
{\hat x} &\equiv& (\cos\theta_2\cos\phi_2, \cos\theta_2 \sin\phi_2, -\sin\theta_2),  \nonumber
\\
{\hat y} &\equiv& (-\sin\phi_2, \cos\phi_2, 0),  \nonumber
\\ 
{\hat z} &\equiv& (\sin\theta_2\cos\phi_2, \sin\theta_2 \sin\phi_2, \cos\theta_2).  \label{hfan}
\end{eqnarray}

Our notation differs from the usual one. In particular, as regards the latter kind of decay, the current literature considers the azimuthal angle, usually denoted  by $\chi$, between the planes ($\vec{p}_2$, $\vec{p}_3$) and (-$\vec{p}_2$, $\vec{p}_l$) in the $R_1$ RF. Taking into account that our azimuthal angles $\phi_3$ and $\phi_l$ refer to   
{\it helicity} RF with opposite $z$-axes ($\pm\vec{p}_2$), one has 
$\chi = \phi_3+\phi_l$: see also Eq. (\ref{azmh}) below, where $\chi$ is identified with $\phi$.

The expressions, that we deduce in Sects. 3 to 6, are consistent with the results of the previous authors; some of them consider a multiple differential decay width\cite{ivn,cohe,cdfz,adm,bkt,gu8,huea}, but generally they integrate over $d\Omega_l$ or over $d\Omega_2$ and, in the case when the decay of $R_2$ is observed, also over $d\Omega_3$. 
 
\subsection{Decay Amplitudes}

We split the decay process into two steps, $R_1 \to R_2 {\cal B}^*$ and ${\cal B}^*\to l \bar{\nu}_l$, where ${\cal B}^*$ is an intermediate boson, which corresponds to $W^*$ in the SM. There are 5 possible kinds of interactions according to Dirac's algebra, $L=V-A$, $R=V+A$, $H=S-P$, $S+P$ and $T$. The matrix element reads, for a given interaction, as
\begin{equation}
{\cal M}_M^{fi}= \sum_{J=0}^1 \sum_\mu [\sqrt{\frac{2S+1}{4\pi}} a^{S,J}_{\lambda\mu}(Q){\cal D}^S_{M\Lambda}(\Omega_2)]
[\sqrt{\frac{2J+1}{4\pi}}c^J_{\nu}(Q){\cal D}^J_{\mu\nu}(\Omega_l)].
\label{oamp}
\end{equation}
Here $a^{S,J}_{\lambda\mu}$ and $c^J_{\nu}$ are the helicity decay amplitudes for each step; $S$ is the spin of $R_1$, $J$ = 0 or 1 the spin of ${\cal B}^*$, $\lambda$ and $\mu$ are the helicities of $R_2$ and ${\cal B}^*$ respectively, $\nu$ = $\lambda_l-1/2$ = 0 or -1 and $\lambda_l$ the helicity of $l$, as we assume a massless antineutrino; $M$ is the $z$-component of the spin of $R_1$ in the canonical RF of the parent resonance and $\Lambda$ = $\lambda-\mu$, while the $\cal D$ are the Wigner rotation functions.

Note that no factorization occurs between the two successive decay
amplitudes, as the sum (\ref{oamp}) runs over 0 and 1, even in the SM, owing to the scalar component of the intermediate virtual boson. This is a consequence of the short lifetime of the intermediate boson. The lack of factorization is made more evident by the presence - if any -  of interactions other than $L$; indeed, one has to replace   
\begin{equation}
a^{S,J}_{\lambda\mu} c^J_{\nu} \to \sum_r a^{S,J,r}_{\lambda\mu} c^{J,r}_{\nu},  \label{rplc}
\end{equation}
where $r$ runs over the 5  possible interactions. Moreover, while the amplitudes $a^{S,J,r}_{\lambda\mu}$ depend on the specific hadron ($B$, $\Lambda_b$, {\it etc.}) and on the model that we choose for describing the form factor, $c^{J,r}_{\nu}$ can be calculated  exactly, as we shall do in Sect. 6.

\section{The $\ell-s$ Scheme}

This scheme\cite{kosc} is useful for describing some features of the three semi-leptonic decays that we have considered. The intermediate, virtual boson ${\cal B}^*$ may carry spin 0 or 1, with positive or negative parity:
\begin{equation}
{\cal B}^*_{V/A}\to (1^{\mp},0^{\mp}), ~~ {\cal B}^*_{S/P}\to 0^{\pm}, ~~ {\cal B}^*_T\to (1^+,1^-),
\end{equation}
where the double assignment in the cases $V$, $A$ and $T$ is due to the virtual character of the boson.

The scheme yields the following results.

\vspace{10pt}
A) $B\to D l \bar{\nu}_l$:
\vspace{10pt}

$B(0^-) \to D(0^-){\cal B}^*(1^-)$: \ ~~~~ \ $P$-wave; \ ~~~~ \ $V,
~ T_V$.  

$B(0^-) \to D(0^-){\cal B}^*(0^+)$: \ ~~~~ \ $S$-wave; \ ~~~~ \ $S, ~  
    A_S$.

\vspace{10pt} 
B) $B\to D^* l \bar{\nu}_l$:
\vspace{10pt}

$B(0^-) \to D^*(1^-){\cal B}^*(1^-)$: \ ~~~~ \ $P$-wave; \ ~~~~ \ $V, ~ T_V$.

$B(0^-) \to D^*(1^-){\cal B}^*(1^+)$: \ ~~~~ \ $S,D$-wave; \ ~  $A, ~ T_A$.

$B(0^-) \to D^*(1^-){\cal B}^*(0^-)$: \ ~~~~ \ $P$-wave; \ ~~~~ \ $P, ~ V_P$.

\vspace{10pt}
C) $\Lambda_b \to \Lambda_c l \bar{\nu}_l$:
\vspace{10pt}

$\Lambda_b(1/2)^+ \to \Lambda_c(1/2)^+ {\cal B}^*(1^-)$: \ ~~~~ \ $s=1/2$, ~ $P$-wave; \ ~~~~ \ $V, ~ T_V$.

$\Lambda_b(1/2)^+ \to \Lambda_c(1/2)^+ {\cal B}^*(1^-)$: \ ~~~~ \ $s=3/2$, ~ $P$-wave; \ ~~~~ \ $V, ~ T_V$

$\Lambda_b(1/2)^+ \to \Lambda_c(1/2)^+ {\cal B}^*(1^-)$: \ ~~~~ \  $s=1/2$, ~ $S$-wave; \ ~~~~ \ $A, ~ T_A$.

$\Lambda_b(1/2)^+ \to \Lambda_c(1/2)^+ {\cal B}^*(1^-)$: \ ~~~~ \  $s=3/2$ ~ $D$-wave;  \ ~~~~ \  $A, ~ T_A$.
 
$\Lambda_b(1/2)^+ \to \Lambda_c(1/2)^+ {\cal B}^*(0^-)$:  \  ~~~~ \ $s=1/2$, ~ $P$-wave; \ ~~~~ \ $P, ~ V_P$.

$\Lambda_b(1/2)^+ \to \Lambda_c(1/2)^+ {\cal B}^*(0^+)$: \ ~~~~ \ $s=1/2$, ~ $S$-wave; \ ~~~~ \ $S, ~ A_S$.
\vspace{10pt}

Here $s$ is the spin of the ${\cal B}^*-\Lambda_c$ system, $T_V$ and $T_A$ denote the contributions of tensor (vector and axial) bosons, while $A_S$ and $V_P$ indicate, respectively, the (pseudo)-scalar components of the axial and vector boson. 

The prospect helps interpreting qualitatively the results by Tanaka and Watanabe\cite{tw} (TW)  about the $B\to D^{(*)} \tau \bar{\nu}_{\tau}$ decays, which are shown in the diagrams of Fig. 1 of their paper. (See also refs. [11,18,87,88,93]). 

- Consider, first of all, the $L=V-A$ and $R=V+A$ currents. Only $V$ contributes in the $B\to D$ transition, which explains why the diagrams corresponding to $L$ and $R$ are equal in this case. On the contrary, the $V$-term is quite small, but not zero, in the $B\to D^*$ transition, which implies that the $R$ contribution is slightly greater than the $L$ one and this latter has a - sign; all that can be established from that figure. 

- As regards the $H=S-P$ contribution, $P$ is forbidden in the $B\to D$ transition, while $S$ does not contribute to the $B\to D^*$ transition, therefore, due to the smallness of the matrix element $P$, a large NP coupling constant is demanded. This is confirmed by fig. 3 of TW, where it also appears that the combination $S+P$ is extremely unlikely. 

- One can also conclude that $T_V$ interferes positively with $V$, while $T_A$ interferes negatively with $A$ and that $|T_A|$ $>$ $|T_V|$\cite{ga}. A similar result will be found in Sect. 6 about the $\Lambda_b$ decay. 

Furthermore, the prospect above shows that the $\Lambda_b \to \Lambda_c$ transition includes the partial waves of both the $B \to D$ and $B \to D^*$ transitions, allowing for interference between them and offering more detailed information. We shall see that also in the helicity representation.
  
\section{Semi-leptonic $B$ Decays}
 
In recent years, the angular distributions of semi-leptonic decays were studied by various authors, mainly as regards the $B\to D^*$ transitions. Several angular observables are singled out, some of which are especially sensitive to NP\cite{bb,bcfn,cdfz,btth,btt2}. Alternatively, a set of such observables are proposed, which, in the framework of the SM, are related to one another: the failure of such relations would imply NP\cite{gro,adm}.

In the present section, we deduce the formulae for the angular distributions of the semi-leptonic $B$ decays and propose few alternative tests for NP. The momentum distribution of the charged lepton is often required, then such formulae are indicated for light leptons; but, in the case of $B\to D^* \tau \bar{\nu}_\tau$ decay, if the secondary decay of the vector meson is observed, information can be inferred also without determining the momentum of $\tau$. Otherwise, the study of a secondary, hadronic $\tau$ decay or of the $\tau$ polarization is proposed\cite{gjl,che,huea,ivn,btt2,As2,Jan,ivn2,aln,celi}.

\subsection{$B\to D l \bar{\nu}_l$}

In this case, we have to insert into Eqs. (\ref{dga})

\begin{equation}
T_D = |a^{0,0}_{00}|^2|c^0_0|^2 + |a^{0,1}_{00}|^2(|c^1_0|^2 \cos^2 \theta_l + \frac{1}{2}|c^1_{-1}|^2 \sin^2 \theta_l) + 2\Re (a^{0,1}_{00}c^1_0 a^{0,0*}_{00}c^{0*}_0)\cos \theta_l,
\label{BDd}
\end{equation}
where the substitution (\ref{rplc}) has to be made if more interactions are considered. Therefore, the angular distribution for the decay reads as	  
\begin{equation}
I(\theta_l,Q) = \frac{1}{2(A_D+C_D/3)}(A_D+B_D\cos \theta_l + C_D\cos^2 \theta_l), \label{and}
\end{equation}
with 
\begin{eqnarray}
A_D &=& |a^{0,0}_{00}|^2|c^0_0|^2 +1/2|a^{0,1}_{00}|^2|c^1_{-1}|^2, ~~ B_D = 2\Re (a^{0,1}_{00}c^1_0 a^{0,0*}_{00}c^{0*}_0), \nonumber
\\
C_D &=& |a^{0,1}_{00}|^2(|c^1_0|^2-1/2|c^1_{-1}|^2). \ ~~~~ \ \ ~~~~ \ \ ~~~~ \ \ ~~~~ \  \ ~~~~ \ \ ~~~~ \ \ ~~~~ \ \ ~~~~ \ \label{abc}
\end{eqnarray}
Here, $I(\theta_l,Q)$ is a shorthand notation for
\begin{equation}
I(\theta_l,Q) = \frac{1}{\Gamma} \frac{d^2\Gamma}{d\cos \theta_l dQ}
\end{equation}
and analogous notations will be adopted in the following. If we integrate this observable over $Q$, we get
\begin{equation}
I(\theta_l) = \frac{1}{2(\bar{A}_D+\bar{C}_D/3)}(\bar{A}_D+\bar{B}_D\cos \theta_l + \bar{C}_D\cos^2 \theta_l), \label{and1}
\end{equation}
with
\begin{equation}
\bar{A}_D (\bar{B}_D,\bar{C}_D) = \int_{m_l}^{M_-} dQ p p_l A_D (B_D,C_D) \label{abb}
\end{equation}
and $M_-$ = $M_B-M_D$. 

The parametrizations (\ref{and}) and (\ref{and1}) hold true for any type of interactions, in view of the replacement (\ref{rplc}); they hint at some tests. 

- First of all, we anticipate some results that we shall deduce in Sects. 6 and 7. $C_D$ and $\bar{C}_D$ - the {\it convexity}\cite{gro} parameters - are negative if the NP interaction is $L$, $R$ or $H$, which generalizes a result found by other authors in the case of the SM\cite{ivn2}. On the other hand, $B_D$ and $\bar{B}_D$ are negative in the cases of $L$- and $R$-interaction. They are sensitive to the relative phase $\psi_0$ of the amplitudes $a^{0,1}_{00}$ and $a^{0,0}_{00}$, which, if  non-trivial, implies TRV, since the final-state interactions are negligibly small in a semi-leptonic decay. Incidentally, this TRV test is alternative to the CP asymmetry proposed by other authors\cite{ags,hgw}. 

- Secondly, the distributions (\ref{and}) and (\ref{and1}) depend very mildly on the form factors, since the coefficients (\ref{abc}) and (\ref{abb}), which depend linearly on them through the amplitudes $a^{0,J}_{00}$, appear both at the numerators and at the denominators. In particular, in the Isgur-Wise (IW) approximation\cite{iw}, one has only one form factor, say $\zeta$, which drops out, so that the data can be fitted by using only the phase $\psi_0$ as a free parameter. In turn, this can be adopted as a first step for a more refined fit with different form factors.

Furthermore, under the same approximation, the $L$-interaction predicts\cite{xrh}
\begin{equation}
a^{0,0}_{00} = \zeta \frac{2p M_B}{Q}F_-, ~~~ a^{0,1}_{00} = \zeta \frac{1}{Q}[(M_B^2-M_D^2)F_+ +Q^2F_-], ~~~ F_{\pm}=\frac{M_B\pm M_D}{2\sqrt{M_B M_D}},
\end{equation}
which can be used as a SM test for light leptons.

- Thirdly, in the case of the SM, the ratios\footnote{The amplitudes $c^J_{\nu}$ are real, as we shall see in Sect. 6.}
\begin{equation}
b_D = B_D/(c^1_0 c^0_0) ~~~ \mathrm{and} ~~~ c_D = C_D/(|c^1_0|^2-1/2|c^1_{-1}|^2) \label{lfu}
\end{equation} 
do not depend on the mass of the charged lepton\cite{gro}; therefore, these observables can be used as tests of LFU for light leptons.

- Last, we observe that the distributions (\ref{and}) and (\ref{and1}) hold true for any semi-leptonic decay, like (\ref{bdt}) and (\ref{bpsi}) to (\ref{incl}), if only the $\theta$-distribution of the charged lepton is considered\cite{gro}. The tests that we have just described can be applied to all of such cases.

\subsection{$B\to D^* (\to D \pi) l \bar{\nu}_l$}

For this decay, one has
\begin{equation}
I(\Omega_l,\theta_D;Q) \propto T_{D^*}, \label{Bdcy}
\end{equation} 
where
\begin{equation}
T_{D^*} = |b^1_{00}|^2 [|a^{0,0}_{00}|^2|c^0_0|^2\cos^2 \theta_D +  {\cal F}_0 |c^1_0|^2 +
{\cal  F}_1 |c^1_{-1}|^2 + 2\Psi \cos \theta_D],  \label{Bdcy}
\end{equation}
$b^1_{00}$ is the decay amplitude $D^* \to D \pi$ and 
\begin{eqnarray}
{\cal F}_0 &=& {\cal F}_{0a} + {\cal F}_{0b}, \ ~~~~ {\cal F}_1 = {\cal F}_{1a} + {\cal F}_{1b}, \ ~~~~ \ \ ~~~~ \  
\\
{\cal F}_{0a} &=& |a^{0,1}_{00}|^2 \cos^2\theta_D \cos^2 \theta_l + \frac{1}{4} \sin^2\theta_D \sin^2 \theta_l (|a^{0,1}_{11}|^2 + |a^{0,1}_{-1-1}|^2), \nonumber 
\\
{\cal F}_{0b} &=& \frac{1}{2}\sin^2\theta_D \sin^2 \theta_l \Re(a^{0,1}_{11} a^{0,1*}_{-1-1} e^{2i\phi}) \nonumber \ ~~~~ \ \ ~~~~ \ \ ~~~~ \ \ ~~~~ \ \
\\
&+& \frac{1}{4}\sin 2\theta_D \sin 2\theta_l \Re(a^{0,1}_{00} a^{0,1*}_{11} e^{-i\phi}+ a^{0,1}_{00} a^{0,1*}_{-1-1} 
e^{i\phi}), \nonumber \ ~~~~ \ \ ~~~~ \ \ ~~~~ 
\\
{\cal F}_{1a} &=& \frac{1}{2} [|a^{0,1}_{00}|^2 \cos^2\theta_D \sin^2 \theta_l +\sin^2\theta_D (|a^{0,1}_{11}|^2
\cos^4 \theta_l/2) + |a^{0,1}_{-1-1}|^2\sin^4 \theta_l/2], \nonumber \ ~~~~ \
\\
{\cal F}_{1b} &=& \frac{1}{2}\{-\frac{1}{2}\sin^2\theta_D \sin^2 \theta_l \Re(a^{0,1}_{11} a^{0,1*}_{-1-1}e^{2i\phi}) \nonumber 
\\ 
&+& \sin 2\theta_D \sin \theta_l [\Re(a^{0,1}_{00} a^{0,1*}_{-1-1}e^{i\phi}) \sin^2 \theta_l/2 - \nonumber 
\\
&-&\Re(a^{0,1}_{00} a^{0,1*}_{11} e^{-i\phi})\cos^2 \theta_l/2]\}, \nonumber \ ~~~~ \
\\
\Psi &=& \Re(c^1_0c^{0*}_0 a^{0,1}_{00} a^{0,0*}_{00})\cos\theta_D \cos\theta_l \nonumber
\ ~~~~ \ \ ~~~~ \ \ ~~~~ \ \ ~~~~ \
\\
&+&\frac{1}{2}\sin\theta_D \sin\theta_l\Re[c^1_0c^{0*}_0 (a^{0,1}_{11}a^{0,0*}_{00} e^{i\phi}+ a^{0,1}_{-1-1} a^{0,0*}_{00} e^{-i\phi})].\label{azmh} \ ~~~~ \ \ ~~~~ \ 
\end{eqnarray}
Here $\phi$ = $\phi_D+\phi_l$; moreover there are 4 independent amplitudes for the decay $ B\to D^*{\cal B}^*$, 
$a^{0,0}_{00}$, $a^{0,1}_{00}$ and $a^{0,1}_{\pm1\pm1}$, corresponding to the 4 form factors for this decay\cite{ivn2,ivn3}. 
The angular distribution that is generated by the amplitude (\ref{Bdcy}) includes interference terms of the type 
\begin{equation}
a^{0,1}_{\xi\xi} a^{0,0*}_{\alpha\alpha} ~~~ \mathrm{or} ~~~ a^{0,1}_{\alpha\alpha}a^{0,1*}_{\beta\beta}~  (\beta\neq\alpha), \label{trv}
\end{equation}
which are sensitive to TRV. In this connection, we stress the importance of the azimuthal dependence of the angular distribution; indeed, in the helicity frame, if a product of the type (\ref{trv}) has an imaginary part, the coefficient of $\sin\phi$ is non-zero, which implies TRV.

A convenient way of searching for TRV consists of determining asymmetries of the type
\begin{equation}
{\cal A} =\frac{N_+ - N_-}{N_+ + N_-}, \label{asmt}
\end{equation}
where $N_{\pm}$ is the number of events for which $\sin \phi$ or $\sin 2\phi$ is positive (negative).

If we integrate Eq. (\ref{Bdcy}) over $d\Omega_l$, we get 
\begin{equation}
I(\theta_D;Q) \propto |b^1_{00}|^2(2 f_L\cos^2\theta_D+f_T \sin^2\theta_D), \label{adst}
\end{equation}
where
\begin{equation}
2f_L=|a^{0,0}_{00}|^2|c^0_0|^2+\frac{2}{3}|a^{0,1}_{00}|^2(|c^1_{-1}|^2+|c^1_0|^2)], \ ~~~~ \   
f_T=\frac{1}{3}(|a^{0,1}_{11}|^2 + |a^{0,1}_{-1-1}|^2)
(|c^1_{-1}|^2+|c^1_0|^2).  \label{ftl}
\end{equation}
In this case, the terms, which could reveal TRV, are washed out.
This can also be seen in a simple way: after integration, ${\cal B}^*$ has a definite helicity -  incidentally, equal to that of $D^*$ - which excludes the possibility of determining any product of the type (\ref{trv}).
By integrating Eq. (\ref{adst}) over $Q$, and suitably normalizing the distribution, we get\cite{ivn2,ivn3}
\begin{equation}
I(\theta_D)=\frac{3}{4}(2F_L^{D^*}\cos^2\theta_D + F_T^{D^*}\sin^2\theta_D), \label{bdc}
\end{equation}
where
\begin{equation}
F_L^{D^*}=\frac{1}{N} \int_{m_l}^{M_-}  dQ p p_l |b^1_{00}|^2 f_L, ~~~ F_T^{D^*}=\frac{1}{N}\int_{m_l}^{M_-}  dQ p p_l |b^1_{00}|^2 f_T \label{fds}
\end{equation}
and 
\begin{equation}
N=\int_{m_l}^{M_-} dQ p p_l |b^1_{00}|^2 (f_L+f_T). \label{fdn}
\end{equation}

We can re-express such variables in terms of the spin density matrix of $D^*$, {\it i. e.} 
\begin{equation}
\rho_{00} = F_L^{D^*}, \ ~~~~ \ \rho_{11}+\rho_{-1-1}=1-\rho_{00}= F_T^{D^*},
\end{equation}
which, owing to angular momentum conservation, equal the analogous variables of ${\cal B}^*$.

It is worth noting that the formulae above hold as well for the decay $B_c\to J/\psi l\nu_l$ and the longitudinal polarization of $J/\psi$ could, in principle, be measured. Moreover, the second Eq. (\ref{ftl})
implies that, according to the SM, the ratio
\begin{equation}
f = f_T/(|c^1_{-1}|^2+|c^1_0|^2)
\end{equation}
does not depend on the mass of the charged lepton and therefore, analogously to the quantities $b_D$ and $c_D$ defined by Eqs. (\ref{lfu}), it can be used as a test of LFU.

\section{Semi-leptonic $\Lambda_b$ Decay}

Now we consider the decay distribution of
$\Lambda_b \to \Lambda_c (\to \Lambda \pi) l \bar{\nu}_l$. According to the second Eq. (\ref{dga}), one has 
\begin{equation}
T_{\Lambda_c} =\sum_{M,M'}\rho^{\Lambda_b}_{MM'} \sum_{\beta\nu} {\cal M}_{M\beta\nu} {\cal M}^*_{M'\beta\nu}, \ ~~~~ \ \rho^{\Lambda_b}_{MM'} = \delta_{MM'}+\vec{\sigma}_{MM'}\cdot \vec{P}. \label{tlac}
\end{equation}
Here $\vec{P}$ is the polarization vector of $\Lambda_b$ and 
\begin{equation}
{\cal M}_{M\beta\nu} \propto b^{1/2}_\beta \sum_{\lambda\mu} {\cal D}^{1/2*}_{M\Lambda} (\Omega_{\Lambda_c}) {\cal D}^{1/2*}_{\lambda\beta} (\Omega_{\Lambda})[a^{1/2,0}_{\lambda~0}c^0_0\delta_{\nu0}
\delta_{\mu0}+ \sqrt{3} a^{1/2,1}_{\lambda~\nu} c^1_\nu {\cal D}^{1*}_{\mu\nu} (\Omega_l)], \label{mdc}
\end{equation}
where $\Lambda$ = $\lambda-\mu$ and $b^{1/2}_\beta$ ($\beta=\pm 1/2$) are the helicity amplitudes for the decay $\Lambda_c \to \Lambda \pi$, such that $|b^{1/2}_+|^2+
|b^{1/2}_-|^2$ =1. Calculations yield
\begin{equation}
T_{\Lambda_c} \propto \Xi_U(\Omega_{\Lambda},\Omega_l)+\Xi_P(\Omega_{\Lambda_c},\Omega_{\Lambda},\Omega_l),
\end{equation} 
where 
\begin{eqnarray}
\Xi_U(\Omega_{\Lambda},\Omega_l) &=& \frac{1}{2}\{A(\theta_l) + \beta_\Lambda B(\theta_l) \cos\theta_{\Lambda}+2\Re[C(\Omega_l)e^{i\phi_{\Lambda}}]\sin\theta_{\Lambda}\}, \nonumber 
\\ 
\Xi_P(\Omega_{\Lambda_c},\Omega_{\Lambda},\Omega_l) &=& \frac{1}{2}{\cal P}_{++} (\Omega_{\Lambda_c})\{D(\theta_l) + \beta_\Lambda E(\theta_l)\cos\theta_{\Lambda} + \nonumber
\\
&+&2\beta_\Lambda \Re[F(\Omega_l)e^{i\phi_{\Lambda}}]\sin\theta_{\Lambda}\} +
2 \Re \{{\cal P}_{+-}(\Omega_{\Lambda_c})[J(\Omega_l)+ \nonumber
\\
&+& \beta_\Lambda L(\Omega_l) \cos\theta_{\Lambda}+ \beta_\Lambda M(\theta_l)] e^{i\phi_{\Lambda}}\sin\theta_{\Lambda}\}.
\end{eqnarray}

Here we have set  
\begin{equation}
\beta_\Lambda = |b^{1/2}_+|^2-|b^{1/2}_-|^2, ~~~ {\cal P}_{++}= \vec{P}\cdot{\hat z} ~~~ \mathrm{and} ~~~ {\cal P}_{+-}= \vec{P}\cdot({\hat x}+i {\hat y}),
\end{equation}
where the unit vectors ${\hat x}$ {\it etc.} are defined by Eqs. (\ref{hfan}). Moreover, as regards $\Xi_U$, we have
\begin{eqnarray}
A(\theta_l) &=& A_0+ A_1 \cos\theta_l+A_2 \cos^2\theta_l, ~~~ B(\theta_l) = B_0+ B_1 \cos\theta_l+B_2 \cos^2\theta_l,
\\
A_0 &=& G_{0+}+G_{0-}+G_{1+}+G_{1-}, ~~~ B_0 = G_{0+}-G_{0-} +G_{1+}-G_{1-}, \nonumber 
\\
A_1 &=& H_{0+}+H_{0-}+H_{1+}+H_{1-}, ~~~ B_1 = H_{0+}-H_{0-} +H_{1+}-H_{1-}, \nonumber 
\\
A_2 &=& K_{0+}+K_{0-}+K_{1+}+K_{1-}, ~~~ B_2 = K_{0+}-K_{0-} +K_{1+}-K_{1-};
\end{eqnarray}
\begin{eqnarray}
G_{0\pm} &=& |a^{1/2,0}_{\pm~0}|^2 |c^0_0|^2 + \frac{1}{2}|a^{1/2,1}_{\pm~0}|^2 |c^1_{-1}|^2, ~~~  G_{1\pm} = \frac{1}{2} |a^{1/2,1}_{\pm~\pm1}|^2 (|c^1_0|^2 + \frac{1}{2}|c^1_{-1}|^2), \nonumber 
\\
H_{0\pm} &=& 2\Re(c^1_0c^{0*}_0 a^{1/2,1}_{\pm~0} a^{1/2,0*}_{\pm~0}), ~~~ \ ~~~ \ ~~~ \ ~~ \ H_{1\pm} = \mp\frac{1}{2}|a^{1/2,1}_{\pm~\pm1}|^2|c^1_{-1}|^2, \nonumber 
\\
K_{0\pm} &=& |a^{1/2,1}_{\pm~0}|^2(|c^1_0|^2 - \frac{1}{2}|c^1_{-1}|^2), ~~~ \ ~~~ \ K_{1\pm} = -\frac{1}{2} 
|a^{1/2,1}_{\pm~\pm1}|^2 (|c^1_0|^2 - \frac{1}{2}|c^1_{-1}|^2);
\end{eqnarray}
\begin{equation}
C(\Omega_l) = \frac{1}{\sqrt{2}}(C_0+C_1\cos\theta_l)\sin\theta_l e^{i\phi_l},
\end{equation}
\begin{eqnarray}
C_0 &=& -\frac{1}{2}|c^1_{-1}|^2(a^{1/2,1}_{+~1} a^{1/2,1*}_{-~0} + a^{1/2,1}_{+~0} a^{1/2,1*}_{-~-1})+ a^{1/2,0}_{+~0} a^{1/2,1*}_{-~-1} c^0_0 c^{1*}_0 - a^{1/2,1}_{+~1} a^{1/2,0*}_{-~0} c^1_0 c^{0*}_0, \nonumber 
\\
C_1 &=& (a^{1/2,1}_{+~1} a^{1/2,1*}_{-~0}- a^{1/2,1}_{+~0} a^{1/2,1*}_{-~-1}) (\frac{1}{2}|c^1_{-1}|^2-|c^1_0|^2). \ ~~~~~~ \ ~~~~~~ \ ~~~~~~ \ ~~~~~~ \ ~~~~~~ \ ~~~~~~ \ ~~~~~~ \
\end{eqnarray}

On the other hand, the functions introduced in $\Xi_P$ read as
\begin{eqnarray}
D(\theta_l) &=& D_0 + D_1 \cos\theta_l+D_2 \cos^2\theta_l, ~~~ E(\theta_l) = E_0+ E_1 \cos\theta_l+ E_2 \cos^2\theta_l,
\\
D_0 &=& G_{0+}-G_{0-}-G_{1+}+G_{1-}, ~~~ E_0 = G_{0+}+G_{0-}-G_{1+}-G_{1-}, \nonumber 
\\
D_1 &=& H_{0+}-H_{0-}-H_{1+}+H_{1-}, ~~~ E_1 = H_{0+}+H_{0-}-H_{1+}-H_{1-}, \nonumber 
\\
D_2 &=& K_{0+}-K_{0-}-K_{1+}+K_{1-}, ~~~ E_2 = K_{0+}+K_{0-}-K_{1+}-K_{1-};
\end{eqnarray}
\begin{equation}
F(\Omega_l) = \frac{1}{\sqrt{2}}(F_0+F_1\cos\theta_l)\sin\theta_l e^{i\phi_l},
\end{equation}
\begin{eqnarray}
F_0 &=& -\frac{1}{2}|c^1_{-1}|^2(a^{1/2,1}_{+~0} a^{1/2,1*}_{-~-1} - a^{1/2,1}_{+~1} a^{1/2,1*}_{-~0})+ a^{1/2,0}_{+~0} a^{1/2,1*}_{-~-1} c^0_0 c^{1*}_0 + a^{1/2,1}_{+~1} a^{1/2,0*}_{-~0} c^1_0 c^{0*}_0, \nonumber 
\\
F_1 &=& -\frac{1}{2}|c^1_{-1}|^2(a^{1/2,1}_{+~0} a^{1/2,1*}_{-~-1} + a^{1/2,1}_{+~1} a^{1/2,1*}_{-~0}) + (a^{1/2,1}_{+~0} a^{1/2,1*}_{-~-1} + a^{1/2,1}_{+~1} a^{1/2,0*}_{-~0}) |c^1_0|^2; 
\end{eqnarray}

\begin{equation}
J(\Omega_l) = \frac{1}{\sqrt{2}}(J_0+J_1\cos\theta_l)\sin\theta_l e^{-i\phi_l},
\end{equation}
\begin{eqnarray}
J_0 &=& -\frac{1}{2}|c^1_{-1}|^2(a^{1/2,1}_{-~-1} a^{1/2,1*}_{-~0} + a^{1/2,1}_{+~0} a^{1/2,1*}_{+~1})+ a^{1/2,1}_{-~-1} 
a^{1/2,0*}_{-~0} c^1_0 c^{0*}_0 - a^{1/2,0}_{+~0} a^{1/2,1*}_{+~1} c^0_0 c^{1*}_0, \nonumber 
\\
J_1 &=& (a^{1/2,1}_{+~0} a^{1/2,1*}_{+~1}- a^{1/2,1}_{-~-1} a^{1/2,1*}_{-~0})(\frac{1}{2}|c^1_{-1}|^2-|c^1_0|^2); \ ~~~~~~ \ ~~~~~~ \ ~~~~~~ \ ~~~~~~ \ ~~~~~~ \ ~~~~~~ \ ~~~~~~ \
\end{eqnarray}

\begin{equation}
L(\Omega_l) = \frac{1}{\sqrt{2}}(L_0+L_1\cos\theta_l)\sin\theta_l e^{-i\phi_l},
\end{equation}
\begin{eqnarray}
L_0 &=& \frac{1}{2}|c^1_{-1}|^2(a^{1/2,1}_{-~-1} a^{1/2,1*}_{-~0} - a^{1/2,1}_{+~0} a^{1/2,1*}_{+~1})- a^{1/2,1}_{-~-1} 
a^{1/2,0*}_{-~0} c^1_0 c^{0*}_0 - a^{1/2,0}_{+~0} 
a^{1/2,1*}_{+~1} c^0_0 c^{1*}_0, \nonumber 
\\
L_1 &=& (a^{1/2,1}_{+~0} a^{1/2,1*}_{+~1}+ a^{1/2,1}_{-~-1} a^{1/2,1*}_{-~0})(\frac{1}{2}|c^1_{-1}|^2-|c^1_0|^2); \ ~~~~~~ \ ~~~~~~ \ ~~~~~~ \ ~~~~~~ \ ~~~~~~ \ ~~~~~~ \ ~~~~~~ \
\end{eqnarray}

\begin{eqnarray}
M(\theta_l) &=& M_0 + M_1 \cos\theta_l+M_2 \cos^2\theta_l, 
\\
M_0 &=& a^{1/2,0}_{+~0} a^{1/2,0*}_{-~0}|c^0_0|^2+\frac{1}{2}a^{1/2,1}_{+~0} a^{1/2,1*}_{-~0}|c^1_{-1}|^2, \nonumber 
\\
M_1 &=& a^{1/2,0}_{+~0} a^{1/2,1*}_{-~0} c^0_0 c^{1*}_0 + a^{1/2,1}_{+~0} a^{1/2,0*}_{-~0} c^1_0 c^{0*}_0, \nonumber 
\\
M_2 &=& a^{1/2,1}_{+~0} a^{1/2,1*}_{-~0}(|c^1_0|^2-\frac{1}{2}|c^1_{-1}|^2). \label{m2}
\end{eqnarray}

We note that the fully differential decay distribution contains several interference terms, which correspond to products of the type $a^{1/2,J}_{\lambda\mu}a^{1/2,J'*}_{\lambda'\nu}$, with different quantum numbers in the two amplitudes. In the case of light leptons, these could allow to detect possible TRV through the azimuthal dependence, analogously to the products (\ref{trv}). Also here one could consider asymmetries of the type (\ref{asmt}), as already illustrated by Geng {\it et al.}\cite{gjl}.

With this decay, integration over $d\Omega_l$, which is suitable in the case of the $\tau$ lepton, does not eliminate all of such products; indeed, we have 
\begin{equation}
I(\Omega_{\Lambda_c},\Omega_{\Lambda};Q) \propto \bar{\Xi}_U + \bar{\Xi}_P, \label{ILb}
\end{equation}
where
\begin{eqnarray}
\bar{\Xi}_U &=& \frac{1}{2}(\bar{A}_{\Lambda_b} + \beta_\Lambda\bar{B}_{\Lambda_b} \cos\theta_{\Lambda}), ~~~~ \ ~~~~~~ \ ~~~~~~ \ ~~~~~~ \ ~~~~~~ \ ~~~~~~ \
\\
\bar{\Xi}_P &=& \frac{1}{2}[{\cal P}_{++}(\bar{D}_{\Lambda_b} +\beta_\Lambda \bar{E}_{\Lambda_b}\cos\theta_{\Lambda}) + 2\beta_\Lambda\Re({\cal P}_{+-} \bar{M}_{\Lambda_b} \sin\theta_{\Lambda}e^{i\phi_{\Lambda}})] \label{azm}
\end{eqnarray} 
and
\begin{eqnarray}
\bar{A}_{\Lambda_b} &=& A_0+\frac{1}{3}A_2, ~~~ \bar{B}_{\Lambda_b} = B_0+\frac{1}{3}B_2, ~~~ \bar{D}_{\Lambda_b}= D_0+\frac{1}{3}D_2, \nonumber
\\
\bar{E}_{\Lambda_b} &=& E_0+\frac{1}{3}E_2, ~~~ \bar{M}_{\Lambda_b}= M_0+\frac{1}{3}M_2. \ ~~~~~~ \ ~~~~~~
\ ~~~~~~ \ ~~~~~~ \ ~~~~~~ \ ~~~~~~ \ ~~~~~~ 
\end{eqnarray}
The last term of (\ref{azm}) is sensitive to the azimuthal asymmetry and therefore to TRV, as follows from Eqs. (\ref{m2}), which contain products of different helicity amplitudes. Also in this case one can investigate this violation by determining an asymmetry of the type (\ref{asmt}), taking into account, in this case, the sign of $\sin \phi_{\Lambda}$.
 
Alternatively, we observe that the mixed product 
\begin{equation}
{\cal T} = \vec{P}\times\vec{p}_{\Lambda_c}\cdot {\vec{p}_{\Lambda}}
\end{equation}
is T-odd. Therefore, in order to detect a possible TRV, the asymmetry (\ref{asmt}) can be re-defined in such a way that $N_{\pm}$ is the number of events for which ${\cal T}$ is positive (negative). 

By inroducing the appropriate normalization factor into the distribution (\ref{ILb}) and integrating  over $Q$, we get \begin{eqnarray}
I(\Omega_{\Lambda_c},\Omega_{\Lambda}) &=& \frac{1}{16\pi^2} \{1+\beta_\Lambda\beta_L^{\Lambda_c} \cos\theta_{\Lambda}+ P\beta_L^{\Lambda_b}+P\beta_\Lambda [h^{{\cal B}^*}\cos\theta_{\Lambda} \cos\theta_{\Lambda_c} \nonumber
\\
&+& (\beta_T^{\Lambda_c}\cos\phi_{\Lambda}+\beta_N^{\Lambda_c}\sin\phi_{\Lambda}) \sin\theta_{\Lambda}\sin\theta_{\Lambda_c}]\}.
\label{ILT}
\end{eqnarray}
Here $P$ = $|\vec{P}|$ and 
\begin{equation}
\beta^{\Lambda_c}_L = \frac{\hat{B}_{\Lambda_b}}{\hat{A}_{\Lambda_b}}, ~~~ \beta^{\Lambda_b}_L = \frac{\hat{D}_{\Lambda_b}}{\hat{A}_{\Lambda_b}}, ~~~  h^{{\cal B}^*}=
\frac{\hat{E}_{\Lambda_b}}{\hat{A}_{\Lambda_b}}, ~~~ \beta_T^{\Lambda_c} = -2\frac{\Re\hat{M}_{\Lambda_b}}{\hat{A}_{\Lambda_b}}, ~~~ \beta_N^{\Lambda_c} = 2\frac{\Im\hat{M}_{\Lambda_b}}{\hat{A}_{\Lambda_b}}, \label{obs1}
\end{equation}
with
\begin{equation}
\hat{A}_{\Lambda_b} (\hat{B}_{\Lambda_b},\hat{D}_{\Lambda_b},\hat{E}_{\Lambda_b},\hat{M}_{\Lambda_b}) = \int_{m_l}^{M_-}  dQ p p_l \bar{A}_{\Lambda_b} (\bar{B}_{\Lambda_b},\bar{D}_{\Lambda_b},\bar{E}_{\Lambda_b},\bar{M}_{\Lambda_b}). \label{obs2}
\end{equation}
If non-trivial, the coefficient $\beta_N^{\Lambda_c}$ implies TRV; the  corresponding asymmetry results in
\begin{equation}
{\cal A} = \int\ d\Omega_{\Lambda_c}\int_0^\pi \sin\theta_{\Lambda} d\theta_{\Lambda} [\int_0^\pi-\int_\pi^{2\pi}]d\phi_{\Lambda} I(\Omega_{\Lambda_c},\Omega_{\Lambda}) = \frac{P}{16}\beta_\Lambda \beta_N^{\Lambda_c}. \label{tray}
\end{equation}
Last, the total decay width for the decay reads as
\begin{equation}
\Gamma = \frac{1}{2^4(2\pi)^7} \frac{p_{\Lambda}}{m_{\Lambda_b}m_{\Lambda_c}}\hat{A}_{\Lambda_b}. 
\label{tga}
\end{equation}

\section{Helicity Amplitudes for $ \Lambda_b \to \Lambda_c  l \bar{\nu}_l$}

According to the substitution (\ref{rplc}), the amplitude for the decay $\Lambda_b \to \Lambda_c {\cal B}^*(\to \tau^- \bar{\nu}_{\tau})$ reads as
\begin{equation}
{\cal M} = G'(A_{SM}+ \sum_r z_r A_r), ~~~ z_r = x_r e^{i\psi_r}, ~~~
G' = V_{bc} \frac{G}{\sqrt{2}}. \label{mampl}
\end{equation}
Here $G$ is the Fermi constant, $V_{bc}$ the Cabibbo-Kobayashi-Maskawa matrix element for the quark transition $b \to c$ and $A_{SM}$ and $z_r A_r$ take account of, respectively, the SM and NP interactions, with $x_r$ real and positive. Now we express the different amplitudes $A$ - defined up to the factor 
$G'$ - in terms of the Dirac operators. 

a) The SM amplitude reads as
\begin{equation}
A_{SM} = A_L = h^\alpha_L g_{\alpha\beta} l^{L\beta}, \label{smal}
\end{equation}
where $h^\alpha_L$ and $l_\alpha^L$ describe, respectively, the hadronic and leptonic vertex, {\it i. e.}, 
\begin{equation}
h^\alpha_L = {\bar u}_{\Lambda_c} (\Gamma^{\alpha}-\Gamma^{\alpha}_5) u_{\Lambda_b} ~~~ \mathrm{and} ~~~ l_\alpha^L = {\bar u}_l \gamma_{\alpha} v_{\bar{\nu}}. \label{SM}
\end{equation}
Here
\begin{equation}
\Gamma^{\alpha} = {\cal F}_1\gamma^{\alpha}+{\cal F}_2i\sigma^{\alpha\beta}q_{\beta} +{\cal F}_3q^{\alpha},
\ ~~~~ \
\Gamma^{\alpha}_5= [{\cal G}_1\gamma^{\alpha}+{\cal G}_2 i\sigma^{\alpha\beta}q_{\beta} +{\cal G}_3 q^{\alpha}]\gamma_5 \label{pscl}
\end{equation}
and the form factors ${\cal F}_i$ and ${\cal G}_i$ are functions of $Q^2$.

b) The $L$-interaction amounts to re-scaling the SM amplitude, so that
\begin{equation}
A_1 = A_L  
\end{equation}
and $z_1$ in Eq. (\ref{mampl}) is a real positive quantity.

c) The $R$-interaction is obtained from the SM expression by changing the sign of $\Gamma^{\alpha}_5$ in the first Eq. (\ref{SM}):
\begin{equation}
A_2 = A_R = h^\alpha_R g_{\alpha\beta}l^{L\beta}, ~~~~ h^\alpha_R =  {\bar u}_{\Lambda_c} (\Gamma^{\alpha} + \Gamma^{\alpha}_5). 
\end{equation}

d) The two combinations of the scalar and pseudoscalar interactions ($S \mp P$) read as 
\begin{equation}
A_{3(4)} = A_H(A_{S+P}) = {\bar u}_{\Lambda_c} ({\cal F}_0 \mp \gamma_5 {\cal G}_0) u_{\Lambda_b}{\bar u}_l v_{\bar{\nu}}, \label{dda}
\end{equation}
where ${\cal F}_0$ and ${\cal G}_0$ are the scalar form factors. These can be related to the previous form factors thanks to the equations of motion\cite{dsfa}: 
\begin{equation}
{\cal F}_0 = {\cal F}_1\rho_- + {\cal F}_3\frac{Q^2}{m_b-m_c}, ~~~~
{\cal G}_0 = {\cal G}_1\rho_+ - {\cal G}_3\frac{Q^2}{m_b+m_c};
\end{equation}
Here 
\begin{equation}
\rho_{\pm} = \frac{m_{\Lambda_b} \pm m_{\Lambda_c}}{m_b \pm m_c} 
\end{equation}
and $m_b$ and $m_c$ are the masses of the $b$- and $c$-quark respectively, $m_b$ = 4.18 $GeV$ and $m_c$ = 1.28 $GeV$.

e) Last, as regards the tensor interaction, we adopt the Isgur-Wise model, according to ref. \cite{bliw}. Then  
\begin{equation}
A_5 = A_T = h_T^{\alpha\beta}g_{\alpha\mu}g_{\beta\nu}l_T^{\mu\nu}, ~~~~ h_T^{\alpha\beta}=\zeta {\bar u}_{\Lambda_c} \sigma^{\alpha\beta} u_{\Lambda_b}, 
~~~~ l_T^{\mu\nu} = {\bar u}_l \sigma_T^{\mu\nu} v_{\bar{\nu}}, \label{tnsr}
\end{equation}
where $\zeta$ is the Isgur-Wise form factor. A more complete parametrization of this amplitude is given in ref. \cite{dt2}.

In the next subsections, we shall derive the expressions of the helicity amplitudes $a^{1/2,J}_{\lambda\mu}$ and $c^J_{\nu}$, defined in Eq. (\ref{oamp}). To this end, we specify $a^{1/2,J}_{\lambda\mu}$ according to Eq. (\ref{mampl}), {\it i. e.},  
\begin{equation}
a^{1/2,J}_{\lambda\mu} = G'({\bar a}^{1/2,J}_{\lambda\mu}+\sum_r{\hat a}^{1/2,J,r}_{\lambda\mu}), \label{hamp}
\end{equation}
where ${\bar a}$ and ${\hat a}^r$ are, respectively, the SM and the NP contributions. However, in order to simplify the notations, we shall drop in the following the 'bar', the 'hat' and the index $r$; therefore, from now on, the symbols $a^{1/2,J}_{\lambda\mu}$ will assume a different meaning than in Eq. (\ref{mdc}).

\subsection{$L$- and $R$-Interaction}

In this case, we introduce four mutually orthogonal unit four-vectors in the Minkowski space-time, {\it i. e.}, $\epsilon^{(m)}_\alpha$, with $m$ = $t,0,\pm 1$ and 
\begin{equation}
\epsilon^{(m)\dagger}\cdot\epsilon^{(m')} = 0 ~~ \mathrm{for} ~~ m'\neq m, ~~ \epsilon^{(t)\dagger}\cdot\epsilon^{(t)} = 1 ~~ \mathrm{and} ~~
\epsilon^{(m)\dagger}\cdot\epsilon^{(m)}= -1 ~~ \mathrm{for} ~~ m\neq t. 
\end{equation}
We exploit the completeness relation for such four-vectors,
\begin{equation}
\epsilon^{(t)}_\alpha\epsilon^{(t)\dagger}_\beta - \epsilon^{(0)}_\alpha\epsilon^{(0)\dagger}_\beta -
\epsilon^{(1)}_\alpha\epsilon^{(1)\dagger}_\beta -
\epsilon^{(-1)}_\alpha\epsilon^{(-1)\dagger}_\beta = g_{\alpha\beta}, \label{cmplt}
\end{equation}
which we insert into Eq. (\ref{smal}). As a result, we get
\begin{equation}
A_L = H_t^L L_L^t - H_0^L L_L^0 - H_1^L L_L^1 - H_{-1}^L L_L^{-1}, \label{hac}
\end{equation}
where 
\begin{equation}
H_{\rho}^L = h^\alpha_L \epsilon^{(\rho)}_\alpha, ~~~ L^{\rho}_L = \epsilon^{({\rho})\dagger}_\alpha l^{L\alpha}, ~~~ \rho=t,0,\pm 1.
\end{equation}
Both $H_{\rho}^L$ and $L^{\rho}_L$ are Lorentz invariant amplitudes, that will be calculated in suitable RF. Similar approaches are proposed in refs. \cite{hgw,hu0,dt2,cht}. These amplitudes will be denoted, from now on, as $H_{\rho\lambda}^L$ and $L^{\rho}_{\lambda_l}$, since they depend also on the helicities of, respectively, $\Lambda_c$ and $l$.

\subsubsection{Hadronic Amplitudes}

We calculate $H_{\rho\lambda}^L$ in a $\Lambda_b$ RF, where $\rho$ and $\lambda$ correspond, respectively, to the helicity of ${\cal B}^*$ and to the helicity of $\Lambda_c$. The unit four-vectors are conveniently expressed by fixing the $z$-axis along $\vec{p}_{\Lambda_c}$. Then, the four-momentum of the $\tau-\bar{\nu}_{\tau}$ system reads as $q\equiv (p_0,\vec{0}_{\perp},-p)$, with $p_0=\sqrt{Q^2+p^2}$; moreover\cite{dsa2}
\begin{equation*} 
\epsilon^{(t)} \equiv\frac{1}{Q}
\begin{pmatrix}
p_0 \\
0 \\
0 \\
-p
\end{pmatrix}, ~~~~~
\epsilon^{(0)} \equiv\frac{1}{Q}
\begin{pmatrix}
-p \\ 
0 \\
0 \\
p_0
\end{pmatrix}, ~~~~~ 
\epsilon^{(\pm1)} \equiv \frac{1}{\sqrt{2}} 
\begin{pmatrix}
0 \\
\mp1 \\
-i \\ 
0
\end{pmatrix}.
\end{equation*}
On the other hand, for our calculation, the quantization axis of $\Lambda_b$ (canonical RF) is suitably chosen along the polarization vector $\vec{P}$. Then, denoting by $\Omega'_{\Lambda_c}$ the direction of $\vec{p}_{\Lambda_c}$ with respect to this RF,
the hadronic amplitudes result in
\begin{eqnarray}
H_{t\lambda}^L &=& \frac{F}{Q} \{{\cal F}_1 (p_0+\chi p) 
+ {\cal F}_3 Q^2- 2\lambda[{\cal G}_1 (p+\chi p_0)-{\cal G}_3\chi Q^2]\} {\cal D}^{1/2*}_{M\lambda} (\Omega'_{\Lambda_c}), \nonumber
\\
H_{0\lambda}^L &=& -\frac{F}{Q}\{{\cal F}_1(p+\chi p_0) +
{\cal F}_2\chi Q^2 -2\lambda[{\cal G}_1(p_0+\chi p) - {\cal G}_2 Q^2]\} {\cal D}^{1/2*}_{M\lambda} (\Omega'_{\Lambda_c}), \nonumber
\\
H_{\pm 1\pm}^L &=& \sqrt{2}F\{[{\cal F}_1\chi - {\cal F}_2 (p+\chi p_0)] \mp[{\cal G}_1-{\cal G}_2(p_0+\chi p)]\} {\cal D}^{1/2*}_{M\mp} (\Omega'_{\Lambda_c}), \label{tal3}
\end{eqnarray}
where, 
\begin{equation}
F = \sqrt{2m_{\Lambda_b}(E_{\Lambda_c}+m_{\Lambda_c})}, ~~~ \chi = \frac{p}{E_{\Lambda_c}+m_{\Lambda_c}}, ~~~ E_{\Lambda_c} = \sqrt{p^2+m_{\Lambda_c}^2}.
\end{equation}
Note that the coefficients of the ${\cal D}$-functions correspond to the hadronic helicity amplitudes up to a sign, which we shall determine below.

\subsubsection{Leptonic Amplitudes}

The amplitudes $L^{\rho}_{\lambda_l}$ describe the decay ${\cal B}^*\to l \nu_l$. We calculate them in a RF of ${\cal B}^*$. But Eq. (\ref{hac}) implies that the spin component of ${\cal B}^*$ along $-\vec{p}_{\Lambda_c}$ is $\rho$. Therefore, in order to calculate the helicity amplitudes for the leptonic vertex, we start from  the ${\cal B}^*$ RF with the quantization axis along $\vec{p}_l$, then we perform a rotation to a frame, whose quantization axis is along $-\vec{p}_{\Lambda_c}$. In the former frame, the conjugated unit four-vectors read as
\begin{equation}
\epsilon^{(t)\dagger} \equiv (1,0,0,0), ~~~ \epsilon^{(0)\dagger} \equiv (0,0,0,1), ~~~ \epsilon^{(\pm1)\dagger} \equiv \frac{1}{\sqrt{2}} (0,\mp1,i,0) 
\end{equation}
and
\begin{equation}
{\hat L}^{t\lambda_l}_L = {\hat L}^{0\lambda_l}_L = F_l(1-\chi_l)\delta_{\lambda_l+}, ~~~
{\hat L}^{1\lambda_l}_L = -F_l (1+\chi_l)\delta_{\lambda_l-}, ~~~ 
{\hat L}^{-1\lambda_l}_L=0,   
\end{equation}
where
\begin{equation}
F_l = \sqrt{p_l(E_l+m_l)}, ~~~ \chi_l = \frac{p_l}{E_l +m_l}, ~~~
E_l = \sqrt{p_l^2+m_l^2}.
\end{equation}
The successive rotation yields
\begin{eqnarray} 
L^{t\lambda_l}_L &=& {\hat L}^{t\lambda_l}_L = c^{0L}_0\delta_{\lambda_l+}, ~~~~~~ \ ~~~~~~ \nonumber 
\\
L^{0\lambda_l}_L &=& \sqrt{3}[{\cal D}^{1*}_{00} (\Omega_l) c^{1L}_0 \delta_{\lambda_l+} + {\cal D}^{1*}_{0-1}(\Omega_l) c^{1L}_{-1} \delta_{\lambda_l-}], \nonumber 
\\
L^{\pm1\lambda_l}_L &=& \sqrt{3}[{\cal D}^{1*}_{\mp10}(\Omega_l) c^{1L}_0 \delta_{\lambda_l+} + {\cal D}^{1*}_{\mp1-1} (\Omega_l) 
c^{1L}_{-1} \delta_{\lambda_l-}],\label{lf}
 \end{eqnarray}
where
\begin{equation}
c^{0L}_0 = -\sqrt{3}c^{1L}_0 = F_l(1-\chi_l), ~~~~  
\sqrt{3}c^{1L}_{-1} = \sqrt{2} F_l(1+\chi_l). \label{hhc}
\end{equation}

\subsubsection{Helicity Amplitudes}
 
Now we insert Eqs. (\ref{tal3}) and (\ref{lf}) into Eq. (\ref{hac}). By slightly changing the notation, $A_L$ $\to$
$ A_{\lambda\lambda_l}^L$, we get 
\begin{eqnarray}
A_{\lambda +}^L &=&  [a^{1/2,0}_{0\lambda} c^{0L}_0 + a^{1/2,1}_{0\lambda}  \sqrt{3} c^{1L}_0
{\cal D}^{1*}_{00}(\Omega_l)]{\cal D}^{1/2*}_{M\lambda}(\Omega'_{\Lambda_c})  \nonumber
\\
&+& \sqrt{3}[a^{1/2,1}_{+1+} {\cal D}^{1/2*}_{M-} (\Omega'_{\Lambda_c}) 
{\cal D}^{1*}_{+10}(\Omega_l) + a^{1/2,1}_{-1-}{\cal D}^{1/2*}_{M+} (\Omega'_{\Lambda_c}) 
{\cal D}^{1*}_{-10}(\Omega_l)] c^{1L}_0 \label{ovl1}
\end{eqnarray}
and
\begin{eqnarray}
A_{\lambda -}^L &=& \sqrt{3}[a^{1/2,1}_{0\lambda} {\cal D}^{1/2*}_{M\lambda} (\Omega'_{\Lambda_c}) 
{\cal D}^{1*}_{0-1}(\Omega_l)+ \nonumber
\\
&+& a^{1/2,1}_{+1+}{\cal D}^{1/2*}_{M-} (\Omega'_{\Lambda_c}) {\cal D}^{1*}_{+1-1}(\Omega_l)+ 
a^{1/2,1}_{-1-}{\cal D}^{1/2*}_{M+} (\Omega'_{\Lambda_c}) {\cal D}^{1*}_{-1-1}(\Omega_l)] c^{1L}_{-1}. \label{ovl2}
\end{eqnarray}
Here 
\begin{eqnarray}
a^{1/2,0}_{0\lambda} &=& \frac{F}{Q} \{{\cal F}_1 (p_0+\chi p) 
+ {\cal F}_3 Q^2-2\lambda[{\cal G}_1 (p+\chi p_0)-{\cal G}_3\chi Q^2]\}, \nonumber 
\\
a^{1/2,1}_{0\lambda} &=& \frac{F}{Q}\{{\cal F}_1(p+\chi p_0) +
{\cal F}_2\chi Q^2 -2\lambda[{\cal G}_1(p_0+\chi p) - {\cal G}_2 Q^2]\}, \nonumber 
\\
a^{1/2,1}_{\pm1\pm} &=& -\sqrt{2}F\{[{\cal F}_1\chi - {\cal F}_2 (p+\chi p_0)] \mp[{\cal G}_1-{\cal G}_2
(p_0+\chi p)]\}.  \label{hhb}
\end{eqnarray}
According to the usual helicity formalism\cite{jw,ch0}, Eqs. (\ref{hhb}) are identified as the hadronic helicity amplitudes, while the leptonic ones are given by Eqs. (\ref{hhc}).
The change of sign of the last two amplitudes (\ref{hhb}), with respect to the coefficients of the last two Eqs. (\ref{tal3}), corresponds to the change of direction for the quantization axis in passing from the hadronic vertex to the leptonic one.

As far as the $R$-amplitudes $a^{JR}_{\mu\lambda}$ are concerned, their expressions are obtained from Eqs. (\ref{hhb}) by changing the signs of the form factors ${\cal G}_i$.

\subsection{$H$-Interaction}

Eq. (\ref{dda}) implies 
\begin{equation}
T^H_{NP}(\lambda,\lambda_l) = ({\cal F}_0+2\lambda {\cal G}_0\chi) F {\cal D}^{1/2*}_{M\lambda} (\Omega'_{\Lambda_c}) F_l(1+\chi_l)
\end{equation}
and comparison with 
\begin{equation}
T^H_{NP}(\lambda,\lambda_l) = a^{1/2,0}_{0\lambda}{\cal D}^{1/2*}_{M\lambda} (\Omega'_{\Lambda_c}) c^{0H}_0
\end{equation}
yields
\begin{equation}
a^{1/2,0}_{0\lambda} = F({\cal F}_0+2\lambda {\cal G}_0\chi), ~~~~
c^{0H}_0 = F_l(1+\chi_l). \label{Hha}
\end{equation}
As discussed in Sect. 3, we do not consider the combination $S+P$, of which the corresponding contribution could be obtained from the former Eq. (\ref{Hha}) by changing the sign of the second term.  

\subsection{$T$-Interaction}

Here we proceed analogously to the case of $L$- and $R$-interaction. By inserting, again, the completeness relation (\ref{cmplt}) into Eq. (\ref{tnsr}), we get
\begin{equation}
A_T = 2 (H^{10}_T L_T^{10} + H^{1-1}_T L_T^{1-1} + H^{0-1}_T L_T^{0-1} - H^{1t}_T L_T^{1t} - H^{0t}_T L_T^{0t} - H^{-1t}_T L_T^{-1t}), \label{htn}
\end{equation}
where 
\begin{equation}
H^{\rho\sigma}_T = h^{\alpha\beta}_T \epsilon^{(\rho)}_\alpha\epsilon^{(\sigma)}_\beta, ~~~ L^{\rho\sigma}_T = \epsilon^{({\rho})\dagger}_\alpha 
\epsilon^{({\sigma})\dagger}_\beta l^{\alpha\beta}_T, ~~~ \rho,\sigma=t,0,\pm 1.
\end{equation}
The first three terms of Eq. (\ref{htn}) correspond to the contributions of the axial vectors, while the remaining ones concern the polar vectors, which are present only when the particle exchanged is virtual, as occurs in our case.
In developing the calculations, we apply the same method as for the $L$-interaction.

A) As regards $H^{\rho\sigma}_T$, the helicity of ${\cal B}^*$ in a $\Lambda_b$ RF results in $\mu$ = $\rho+\sigma$, $|\mu|\leq 1$. Moreover, initially, we fix the with the $z$-axis along $\vec{p}_{\Lambda_c}$, then we perform a frame rotation to the canonical RF that we have defined above. We get 
\begin{eqnarray}
H^{t0}_T &=& i a^{1/2,1}_{0\lambda,V} {\cal D}^{1/2*}_{M\lambda} (\Omega'_{\Lambda_c}), \ ~~~~ \ \ ~~~~ \ \ ~~~~ \
H^{1-1}_T = i a^{1/2,1}_{0\lambda,A}{\cal D}^{1/2*}_{M\lambda}(\Omega'_{\Lambda_c}), \nonumber
\\  
H^{t\pm 1}_T &=& i a^{1/2,1}_{\pm 1\pm,V}{\cal D}^{1/2*}_{M\mp}(\Omega'_{\Lambda_c}), ~~~~ \ ~~~~ \ ~~~~ \ 
H^{0\pm 1}_T = i a^{1/2,1}_{\pm 1\pm,A}{\cal D}^{1/2*}_{M\mp}(\Omega'_{\Lambda_c}), \label{htn2}
\end{eqnarray}
where
\begin{equation}
a^{1/2,1}_{0\lambda,V} = \zeta\chi F, ~ a^{1/2,1}_{0\lambda,A} = 2\lambda \zeta F, ~ a^{1/2,1}_{\pm 1\pm,V} = \sqrt{2}\zeta F \frac{p+p_0\chi}{Q}, 
 ~ a^{1/2,1}_{\pm 1\pm,A} = \sqrt{2}\zeta F \frac{p_0+p\chi}{Q}. \label{Hth}
\end{equation}

B) In order to calculate the terms $L^{\rho\sigma}_T$, we preliminarily use the helicity frame for $l$ in the ${\cal B}^*$ RF, then we perform a frame rotation from $\vec{p}_l$ to $-\vec{p}_{\Lambda_c}$. The first step yields
\begin{eqnarray}
{\hat L}^{t0}_T &=& i F_l (1+\chi_l)\delta_{\lambda_l +}, ~~~~ {\hat L}^{1-1}_T = -i F_l (1+\chi_l)\delta_{\lambda_l +},  \nonumber 
\\
{\hat L}^{t1}_T &=& {\hat L}^{01}_T = -\sqrt{2}i F_l (1-\chi_l) \delta_{\lambda_l -}, ~~ {\hat L}^{t-1}_T = {\hat L}^{0-1}_T = 0.
\end{eqnarray}
The successive frame rotation gives rise to
\begin{eqnarray}
L^{t0}_T &=& i\sqrt{3} [c^{T_V}_{-1}\delta_{\lambda_l -}{\cal D}^{1*}_{0-1}(\Omega_l) + c^{T_V}_0\delta_{\lambda_l +}{\cal D}^{1*}_{00}(\Omega_l)],  \nonumber 
\\
L^{1-1}_T &=& i\sqrt{3} [c^{T_A}_{-1}\delta_{\lambda_l -}{\cal D}^{1*}_{0-1}(\Omega_l) + c^{T_A}_0\delta_{\lambda_l +}{\cal D}^{1*}_{00}(\Omega_l)], \nonumber 
\\
L^{t\pm 1}_T &=& i\sqrt{3} [c^{T_V}_{-1}\delta_{\lambda_l -}{\cal D}^{1*}_{\pm 1-1}(\Omega_l) + c^{T_V}_0\delta_{\lambda_l +}{\cal D}^{1*}_{\pm 10}(\Omega_l)], \nonumber 
\\ 
L^{0 \pm 1}_T &=& i\sqrt{3} [c^{T_A}_{-1}\delta_{\lambda_l -}{\cal D}^{1*}_{\pm 1-1}(\Omega_l) + c^{T_A}_0\delta_{\lambda_l +}{\cal D}^{1*}_{\pm 10}(\Omega_l)], \label{ltn2}
\end{eqnarray}
where
\begin{equation}
\sqrt{3}c^{T_A}_0 = -\sqrt{3}c^{T_V}_0 = -F_l(1+\chi_l), ~~~~ \sqrt{3}c^{T_A}_{-1} = \sqrt{3}c^{T_V}_{-1} = \sqrt{2} F_l(1-\chi_l). \label{wtn}
\end{equation}
The results (\ref{Hth}) and (\ref{wtn}) lead to two conclusions.

- $|a^{1/2,1}_{\mu\lambda,A}c^{T_A}_\nu|$ $\geq$ $|a^{1/2,1}_{\mu\lambda,V}c^{T_V}_\nu|$  for any $\mu,\lambda,\nu$.

- The $T$-contribution is quite small in comparison with the SM term, except for the cases of $\nu$ = 0 and ($\mu,\lambda$) = ($\pm1,\pm$). Since the phase of the $T$-coupling is small\cite{tw}, the $A-T_A$ interference is positive as well as $V-T_V$ in the case ($1,+$), whereas $A-T_A$ is negative and $V-T_V$ positive in the case ($-1,-$).  

Both conclusions are in agreement with the observations made in Sect. 3 about the $T$-coupling.

C) Inserting Eqs. (\ref{htn2}) and (\ref{ltn2}) into Eq. (\ref{htn}) yields
\begin{eqnarray}
A^T_{\nu} &=& \sqrt{3}[a^{1/2,1}_{+1+}{\cal D}^{1/2*}_{M-} (\Omega'_{\Lambda_c}){\cal D}^{1*}_{1-1}(\Omega_l) +\sum_{\lambda}a^{1/2,1}_{0\lambda} {\cal D}^{1/2*}_{M\lambda} (\Omega'_{\Lambda_c}) {\cal D}^{1*}_{0-1}(\Omega_l) \nonumber
\\
&+& a^{1/2,1}_{-1-}{\cal D}^{1/2*}_{M+} (\Omega'_{\Lambda_c}){\cal D}^{1*}_{-1-1}(\Omega_l)]c^T_{\nu},  \ ~~~~ \ \ ~~~~ \
 \nu = \lambda_l-1/2. \ ~~~~ \ \ ~~~~ \
\end{eqnarray}
Here  the coefficients
\begin{equation}
a^{1/2,1}_{+1+} = a^{1/2,1}_{-1-} = 2\sqrt{2}\zeta\frac{F}{Q}(p+2\lambda_l p_0)(1+2\lambda_l \chi), ~~~ a^{1/2,1}_{0\lambda} = 2\zeta F (\chi+ 4\lambda\lambda_l) 
\end{equation}
and
\begin{equation}
\sqrt{3}c^T_0 = F_l(1+\chi_l),  ~~~ \sqrt{3}c^T_{-1} = \sqrt{2}F_l(1-\chi_l)\label{Tlp}
\end{equation}
are, respectively, the hadronic and the leptonic helicity amplitudes.

\subsection{Remarks}

In Appendix A, the non-covariant expressions of the helicity amplitudes are converted into Lorentz-invariant ones. Moreover, it is worth stressing that the expressions of the coefficients $c^J_{\nu}$, that we have calculated for the various kinds of interactions, are valid independent of the hadrons involved in the semi-leptonic decay that we consider. These results will be used in the next section, in order to deduce important consequences. 

\section{Phenomenology}

Now we make a set of three assumptions, which, as we shall see, imply some important consequences on various semi-leptonic decays. In particular, in the case of the decay $\Lambda_b\to \Lambda_c \tau \bar{\nu}_\tau$, the consequences of these assumptions, together with the
results of the previous sections, lead us to calculating the observables $\beta_L^{\Lambda_c}$, $\beta_T^{\Lambda_c}$,  $\beta_N^{\Lambda_c}$, $\beta_L^{\Lambda_b}$ and $h^{{\cal B}^*}$, which are defined by Eqs. (\ref{obs1})-(\ref{obs2}). 

\subsection{Assumptions}

~~~ - The semi-leptonic decays that involve the $\ell - \bar{\nu}_{\ell}$ system ($\ell$ = $e$, $\mu$) in the final state are consistent with the SM predictions, the possible deviations being very tiny.

- Only one of the possible NP interactions contributes to the anomaly observed in $B \to D^{(*)} \tau \bar{\nu}_{\tau}$ decays, so that the sum (\ref{rplc}) amounts to two terms.\footnote{We shall cite in Sect. 8 several papers where more NP operators are assumed.}

- Such an interaction causes the same anomaly in any process that involves the $\tau - \bar{\nu}_{\tau}$ system in the final state. Quantitatively, given any semi-leptonic decay of that type, whose width is $\Gamma_\tau$, one defines the parameter $\Delta$ as

\begin{equation}
\Gamma_\tau = \Gamma_\tau^{SM}(1+\Delta), \label{asspt}
\end{equation}
which is assumed to be independent of the specific decay where the $\tau - \bar{\nu}_{\tau}$ system is involved. This is an approximation, since the total width depends on the form factor, and therefore on the particular process; however, the ratio of $\Gamma_\tau$ to $\Gamma_\tau^{SM}$ has a very smooth dependence on this quantity: as we shall discuss below, the approximation is acceptable within errors; see also ref. \cite{rdt}. 

Now we deduce two consequences about the two above mentioned semi-leptonic decays of $B$.

A) Assume the NP interaction to be $R$ = $V+A$. As we have seen in Subsect. 2.2, the transition $B \to D$ involves only the $V$-interaction, while $B \to D^*$ is interested almost exclusively by the $A$-term (See also refs. \cite{tw,kosc}). Then the amplitudes read as
\begin{eqnarray}
\mathrm{Ampl}(B \to D \tau \bar{\nu}_{\tau}) &=& V_D(+1+x e^{i\psi}),
\\
\mathrm{Ampl}(B \to D^* \tau \bar{\nu}_{\tau}) &\simeq& A_{D^*}(-1+x e^{i\psi}),
\end{eqnarray}
where $V_D$ and $A_{D^*}$ are SM amplitudes for the processes considered. Then our assumption implies 
\begin{equation}
|+1+x e^{i\psi}|^2 \simeq |-1+x e^{i\psi}|^2 ~~ => ~~ \psi \simeq \pm\frac{\pi}{2},
\end{equation}
which agrees with previous analyses\cite{tw}.

B) Now consider a NP interaction of the type $H$ = $S-P$. In this case, the widths read as
\begin{eqnarray}
\Gamma_D &\propto& |V^1_D|^2 +|V^0_D +S_Dx e^{i\psi}|^2 = {\tilde\Gamma}_D^{SM}+2x\cos\psi V^0_D S_D+x^2 S_D^2,
\\
\Gamma_{D^*} &\propto&  \simeq |A^1_{D^*}|^2+|A^0_{D^*}-P_D^*x e^{i\psi}|^2 = {\tilde\Gamma}_D^{*SM} + 2x\cos\psi A^0_{D^*}P_{D^*} + x^2 P_{D^*}^2,
\end{eqnarray}
where $V^0_D$, $A^0_{D^*}$, $V^1_D$ and $A^1_{D^*}$ are the moduli of the vector and axial SM amplitudes that correspond to spin-0 and spin-1 ${\cal B}^*$ exchange; moreover
\begin{equation}
{\tilde\Gamma}_D^{SM} = |V^0_D|^2 +|V^1_D|^2 ~~ \mathrm{and} ~~ {\tilde\Gamma}_{D^*}^{SM} \simeq |A^0_{D^*}|^2 +|A^1_{D^*}|^2.
\end{equation}
Then it follows from our assumption  
\begin{equation}
(\frac{S_D^2}{\Gamma_D^{SM}}-\frac{P_{D^*}^2}{\Gamma_{D^*}^{SM}})x \simeq -2(\frac{S_D V_D^0}{\Gamma_D^{SM}}+
\frac{P_{D^*}A_{D^*}^0}{\Gamma_{D^*}^{SM}}) \cos\psi;
\end{equation}
since $P_{D^*}$ $<<$ $S_D$\cite{tw}, the coefficient of $x$ is positive, whence we deduce $\cos\psi$ $<$ 0. Also this result is in agreement with the TW analysis. This encourages us to apply our assumption also to the $\Lambda_b$ decay.

\subsection{Predictions on Various Decays}

The $HFLAV$ analysis\cite{hflav} of $B$ semi-leptonic data provides
\begin{eqnarray}
{\cal R}_{D}~ &=& 0.342 \pm 0.026, ~~~~ {\cal R}^{SM}_{D} = 0.298 \pm 0.004 ~~ \to \Delta = 0.148^{+0.104}_{-0.101}, \label{rd}
\\
{\cal R}_{D^*} &=& 0.287\pm 0.012, ~~~~~ {\cal R}^{SM}_{D^*} = 0.254 \pm 0.005 ~~ \to \Delta = 0.130^{+0.071}_{-0.068}. \label{rdstar}
\end{eqnarray}	 
The two values of $\Delta$ are compatible within errors, in agreement with our third assumption. Therefore, we assume for that parameter the weighted average of those indicated in (\ref{rd}) and (\ref{rdstar}):
\begin{equation}
\Delta = 0.136^{+0.063}_{-0.061}. \label{newdlt}
\end{equation}
Then, our assumptions entail
\begin{equation}
\frac{{\cal R}_{\Lambda_c}}{{\cal R}^{SM}_{\Lambda_c}} \simeq 
\frac{{\cal R}_{J/\psi}}{{\cal R}^{SM}_{J/\psi}} \simeq \frac{{\cal R}_{X_c}}{{\cal R}^{SM}_{X_c}} 
\simeq \frac{{\cal R}_L}{{\cal R}^{SM}_L} \simeq 1+\Delta, \label{prdn}
\end{equation}
where $\Delta$ assumes the value (\ref{newdlt}).
 The double ratios in Eq. (\ref{prdn}) refer, respectively, to the semi-leptonic decays $\Lambda_b\to \Lambda_c l \bar{\nu}_l$, $B_c\to J/\psi l \bar{\nu}_l$ and $B\to X_c l \bar{\nu}_l$ and to the leptonic decay $B_c\to l \bar{\nu}_l$. Our first prediction (\ref{prdn}) agrees, within errors, with the results of some more elaborated sum rules\cite{fb}. Similar predictions can be made about other semi-leptonic decays of beauty baryons\cite{shn,rdk}.
 
\subsection{Calculations of $\Gamma^{SM}$ for $\Lambda_b$ Decays - Prediction of ${\cal R}_{\Lambda_c}$}

In order to do further specific predictions on the $\Lambda_b$ decay,
 we determine the total width of a semi-leptonic decay $\Lambda_b\to \Lambda_c l \bar{\nu}_l$, both for the muon and for the $\tau$ lepton. To this end, we use Eq. (\ref{tga}), inserting one of the form factors found in the literature[92,104-112] and the appropriate lepton mass into the expressions of the helicity amplitudes. The parametrizations of the form factors are listed in Appendix B. Table \ref{tab:one} shows, for the various form factors, the SM predictions of $\Gamma_\mu$ and $\Gamma_\tau$ and the ratio ${\cal R}_{\Lambda_c}^{SM}$ = $\Gamma_\tau/\Gamma_\mu$. The symbols $SR_i$ ($i$ = 1,2,3,4), which appear in the table, correspond to the parametrizations listed in Table 1 of ref. \cite{dsfa}. 

\begin{table*}
\begin{center}
\caption{Total width SM predictions for $\Gamma_\mu$ and $\Gamma_\tau$, expressed in $\mu eV$.}
\begin{tabular}{|c|c|c|c|c}
\hline\hline
$~~~~FF~~~~$ & $~~~~Ref.~~~$ & $\Gamma^{SM}_\mu$ & $\Gamma^{SM}_\tau$ &${\cal R}_{\Lambda_c}^{SM}$ \\
\hline\hline
$~~~~IW_1~~~$ & \cite{klw} & 30.36                     &  9.20   & 0.303          \\
$~~~~IW_2~~~$ & \cite{blr} & $23.27^{+2.25}_{-2.13}$   & $7.87^{+0.27}_{-0.30}$  & $0.338^{+0.20}_{-0.19}$ \\
$~~~~IW_3~~~$ & \cite{rhk} & 23.33                     &  7.72   & 0.331          \\
$~~~~Az~~~~~$ & \cite{az} & 89.48                     & 22.55   & 0.252           \\
$~~~~De~~~~~$ & \cite{de} & 21.73                     &  7.26   & 0.334           \\
$~~~~SR_1~~~$ & \cite{dec} & 14.15                     &  4.71   & 0.333          \\
$~~~~SR_2~~~$ & \cite{dec} & 16.96                     &  5.55   & 0.327          \\
$~~~~SR_3~~~$ & \cite{dec} & 27.77                     &  8.36   & 0.301          \\
$~~~~SR_4~~~$ & \cite{dec} & 32.01                     &  9.44   & 0.295          \\
$~~~~Gu~~~~~$ & \cite{gu8} & 45.88                     & 16.06   & 0.350          \\
$~~~~Ke~~~~~$ & \cite{ke}  & 27.14                     &  8.85   & 0.326          \\
$~~~~Li~~~~~$ & \cite{lly} & 27.07                     &  8.45   & 0.312          \\
$~~~~FG~~~~~$ & \cite{fg} & 26.47                     &  8.23   & 0.311           \\
\end{tabular}
\label{tab:one}       
\end{center}
\end{table*}

The experimental value of $\Gamma_\mu$ is\cite{pdg}
\begin{equation}
\Gamma_\mu = (27.74^{+6.48}_{-5.95})~\mu eV 
\label{emu}
\end{equation}
The values of $\Gamma_\mu$, that we have found by means of the form factors $Az$, $SR_1$, $SR_2$ and $Gu$, differ strongly from it, therefore such form factors are discarded in the calculations of the observables. By taking into account the other values reported on Table \ref{tab:one}, the theoretical values of $\Gamma^{SM}_\mu$ and ${\cal R}^{SM}_{\Lambda_c}$ result to be, respectively, 
\begin{equation}
\Gamma^{SM}_\mu = (25.98\pm0.99)~\mu eV ~~~ \mathrm{and} ~~~ {\cal R}^{SM}_{\Lambda_c} = 0.321\pm0.005,
\label{thv}   
\end{equation}
the latter being compatible with the current value\cite{fb}, $0.324\pm0.004$.

Eq. (\ref{prdn}) leads, through the second Eq. (\ref{thv}), to the prediction
\begin{equation}
{\cal R}_{\Lambda_c} = 0.365^{+0.026}_{-0.025},  \label{rtl}
\end{equation}
which agrees, within errors, with the result of a sum rule\cite{fb}: ${\cal R}_{\Lambda_c} = 0.380\pm0.012\pm0.005$.

On the other hand, the experimental value found by the LHCb collaboration\cite{cern} is 
\begin{equation}
{\cal R}_{\Lambda_c}^{exp} = 0.242\pm 0.026 (stat.)\pm 0.040 (syst.) \pm 0.059; \label{gesp}
\end{equation}

This value is compatible with (\ref{rtl}) within errors, while the elimination of possible systematic uncertainties\cite{BLN} is expected to increase the central value of (\ref{gesp}).

\subsection{Fixing the NP Couplings for the Decay $\Lambda_b\to \Lambda_c \tau \bar{\nu}_{\tau}$}

Now we determine the NP coupling, $z$ = $x$ $e^{i\psi}$, taking into account, for each kind of interaction, the bounds established by the analyses on the $B$ decays[54,58,84,86] and, wherever possible, the constraint (\ref{asspt}).

- The $L$-interaction sets no limits on $\psi$, then, choosing it to be zero, we get 
\begin{equation}
x_L = \sqrt{1+\Delta}-1 = 0.066\pm0.029. \label{xl}
\end{equation}

- The $R$-interaction demands, according to the previous analyses \cite{ivn2,tw} and to our deduction, $\psi_R$ = $\pm\pi/2$, whence
\begin{equation}
x_R = \sqrt{\Delta} = 0.369^{+0.077}_{-0.095}. \label{xr}
\end{equation}

- As regards the $H$-interaction, our calculations and our assumptions involve bounds on $x$ and $\psi$ both by the analysis of the semi-leptonic decay of $B$\cite{ivn2} and by our procedure. By considering Eq. (\ref{asspt}), and calculating $\Gamma_\tau$ and $\Gamma_\tau^{SM}$ according to the equations of Sects. 4 to 6, we get the following relation between such parameters: 
\begin{equation}
r_0 x_H^2 + r_1 x_H \cos\psi = \Delta, \label{xpsi}
\end{equation}
where $r_0$ and $r_1$ depend on the specific decay. Our analysis about the $\Lambda_b$ decay yields $r_0$ = 0.784 and $r_1$ = 0.712. Eq. (\ref{xpsi}) corresponds to a circumference in the Argand plane of $z$ = $x_H e^{i\psi_H}$; then, the values of $z_H$ are constrained inside a circular crown, due to the uncertainty on $\Delta$. On the other hand, the $B$-data entail\cite{ivn2} $r_0$ = 0.635 and $r_1$ = 0.965, together with the inequality $x_H \cos\psi_H$ $\leq$ -0.76. Altogether, such bounds imply $\psi$ $\simeq$ $2.274$ rad., in agreement with our predictions above; $x_H$ is the corresponding positive root of Eq. (\ref{xpsi}), with $r_0$ and $r_1$ fixed according to our calculations; it results in 
\begin{equation}
x_H = 0.803^{+0.074}_{-0.083}.
\end{equation}

- Last, the coupling of the $T$-interaction is greatly constrained, owing to the $B$-data\cite{ivn2}, $x_T$ $\sim$ $0.28$ and $\psi_T$ $\sim$ $0.53$ rad., with negligible uncertainties. In this case, we cannot impose a constraint of the type (\ref{xpsi}). Our assumption implies a prediction
\begin{equation}
{\cal R}_{\Lambda_c} = 0.79^{+0.06}_{-0.05},
\end{equation}
quite outside the experimental value (\ref{gesp}). 
    
\subsection{Results about the $\Lambda_b$ Decay to $\Lambda_c \tau \bar{\nu}_{\tau}$}

Now we use Eqs. (\ref{tlac}), (\ref{mdc}) and (\ref{hamp}) to determine the parameters which appear in Eq. (\ref{ILT}). The results of our analysis, which vary according to the NP interaction, are summarized in Table \ref{tab:two}. Eq. (\ref{ILT}) implies that the effective observables, which can be inferred from data of the differential decay distribution, consist of the products 
\begin{equation}
{\cal P}_L = \beta_{\Lambda} \beta_L^{\Lambda_c}, \ ~~~~ \ {\cal P}_{T(N)} = P \beta_{\Lambda} \beta_{T(N)}^{\Lambda_c} \ ~~~~ \ \mathrm{and} \ ~~~~ {\cal H} = P \beta_{\Lambda} h^{{\cal B}^*}. 
\end{equation}
Then, since $\beta_\Lambda$ = $-0.84\pm 0.09$\cite{pdg}, the observable 
${\cal P}_L$ allows to distinguish among the various NP interactions. On the other hand, ${\cal P}_T$, ${\cal P}_N$ and 
${\cal H}$ depend crucially on the polarization $P$ of ${\Lambda_b}$.

In particular, as observed in Sect. 5, the observable $\beta_N^{\Lambda_c}$ is T-odd and therefore it is sensitive to TRV. According to Eq. (\ref{tray}), this gives rise to an asymmetry which, in modulus, amounts to
\begin{equation}
L:R:H:T ~=~ 0.000:0.004P:(0.010^{+0.003}_{-0.002})P:0.003P. \label{tvas}
\end{equation}

On the other hand, $h^{{\cal B}^*}$ allows just to discriminate between $T$ interaction and the other ones. It is related to the longitudinal polarization of the intermediate boson ${\cal B}^*$, 
\begin{equation}
F_L^{{\cal B}^*} = \frac{1}{2}(1+ h^{{\cal B}^*}), 
\end{equation}
which yields
\begin{equation}
~~~~  L,R:H:T ~=~ (0.562\pm0.007):(0.558^{+0.012}_{-0.014}):(0.443\pm0.002). \label{algnm}
\end{equation}
The result will be compared with $F_L^{D^*}$ below.
 
\begin{table*}
\begin{center}
\caption{Predictions of observables according to the different interactions.}
\begin{tabular}{|c|c|c|c|c|c}
\hline\hline
$~~~~~~~~~~$ & $\beta_L^{\Lambda_b}$ & $h^{{\cal B}^*}$ & $\beta_L^{\Lambda_c}$ &$\beta_T^{\Lambda_c}$
&$\beta_N^{\Lambda_c}$ \\
\hline\hline
$~~~L~~~$ & $-0.263\pm 0.021$ &$ 0.124\pm 0.014$ & $-0.659\pm 0.009$ &  $0.104\pm 0.006$   &$ 0.0\pm 0.000$ \\
$~~~R~~~$ & $-0.200\pm 0.025$ &$ 0.124\pm 0.014$ & 
$-0.501^{+0.066}_{-0.061}$ & $0.104\pm 0.006$  & $-0.085\pm 0.003$  \\
$~~~H~~~$ & $-0.245^{+0.027}_{-0.023}$ & $0.115\pm 0.019$   & $-0.443^{+0.053}_{-0.045}$ & $0.140^{+0.019}_{-0.021}$ & $-0.199^{+0.021}_{-0.019}$  \\
$~~~T~~~$ & $-0.345\pm0.003$ & $-0.115\pm 0.003$& $-0.806\pm 0.000$ & $0.063\pm 0.000$ & $-0.054\pm 0.000$ \\
\end{tabular}
\label{tab:two}       
\end{center}
\end{table*}

\subsection{Consequences on the $B\to~D~\tau~\bar{\nu}_\tau$ Angular Distribution}

Now we show that our assumption determines, for some of the NP interactions, the signs of some coefficients of the distributions (\ref{and}) and (\ref{and1}). To this end, we rewrite those coefficients according to the substitution (\ref{rplc}) for the above mentioned interactions. As regards $B_D$, we have

\vspace{10pt}
$L: ~~~~~~~~ B_D ~=~ 2(\bar{a}^{0,1}_{00}+a^1_L) (\bar{a}^{0,0}_{00}+a^0_L)^*c^1_0c^0_0 ~=~ 
-2\bar{a}^{0,1}_{00} \bar{a}^{0,0}_{00}(\sqrt{1+\Delta})^2 F_l^2 (1-\chi_l^2) ~ < ~ 0;$
 
$R: ~~~~~~~~ B_D ~=~ 2(\bar{a}^{0,1}_{00}+a^1_R) (\bar{a}^{0,0}_{00}+a^0_R)^*c^1_0c^0_0 ~=~ 
-2\bar{a}^{0,1}_{00} \bar{a}^{0,0}_{00} |1\pm i\sqrt{\Delta}|^2 F_l^2 (1-\chi_l^2) ~ < ~ 0.$
\vspace{10pt}

Here, the barred symbols refer to the SM amplitudes and $a^{0(1)}_{L(R)}$ are the NP contributions; moreover, Eqs. (\ref{hhc}), (\ref{xl}) and  (\ref{xr}) have been taken into account. Analogously, for $C_D$,

\vspace{10pt}
$L: ~~~~~~~~ C_D ~=~ |\bar{a}^{0,1}_{00}+a^1_L|^2 [(c^1_0)^2-\frac{1}{2}(c^1_{-1})^2] ~=~ |\bar{a}^{0,1}_{00}|^2 (1+\Delta) F_l^2 (-4\chi_l) ~ < ~ 0;$

$R: ~~~~~~~~ C_D ~=~ |\bar{a}^{0,1}_{00}+a^1_R|^2 [(c^1_0)^2-\frac{1}{2}(c^1_{-1})^2] ~=~ |\bar{a}^{0,1}_{00}|^2 (1+\Delta) F_l^2 (-4\chi_l) ~ < ~ 0;$

$H: ~~~~~~~~ C_D ~=~ |\bar{a}^{0,1}_{00}|^2 [(c^1_0)^2-\frac{1}{2}(c^1_{-1})^2] ~~=~~ |a^{0,1}_{00}|^2  F_l^2 
(-4\chi_l) ~ < ~ 0.$ 
\vspace{10pt}

Similar conclusions are drawn for $\bar{B}_D$ and $\bar{C}_D$. All that proves what we asserted in Subsect. 4.1.

\subsection{The observable $F_L^{D^*}$}

We show the predictions about $F_L^{D^*}$ according to the various NP interactions\cite{kkr}. Recalling Eq. (\ref{fdn}), we set 
\begin{equation}
N = \bar{N} (1+\Delta),
\end{equation}
where the barred quantity refers, as before and from now on, to the SM contribution. Taking into account Eqs. (\ref{fds}), we get

- for the $L$- and $R$-interaction,  
\begin{equation}
N F_L^{D^*} = \bar{N} \bar{F}_L^{D^*} (1+\Delta) \implies  F_L^{D^*} = \bar{F}_L^{D^*}; \label{lri}
\end{equation}

- for the $H$-interaction,
\begin{equation}
N F_L^{D^*} = \bar{N} \bar{F}_L^{D^*}+\bar{N}\Delta \implies F_L^{D^*} = 
\frac{\bar{F}_L^{D^*}+\Delta}{1+\Delta}; \label{hig}
\end{equation}
 
 - for the $T$-interaction, starting from $F_T^{D^*}$ and reasoning analogously to Eq. (\ref{hig}), 
\begin{equation}
F_T^{D^*} = \frac{\bar{F}_T^{D^*}+\Delta}{1+\Delta} \implies  F_L^{D^*} = \frac{\bar{F}_L^{D^*}}{1+\Delta}.
\end{equation}

Then, taking into account the SM predictions of the observable\cite{btth,hug} and the fits to the recent measurements of $F_L^{D^*}$ for light leptons\cite{fb,fdl24}, we get the following results:
\begin{eqnarray}
\mathrm{Ref.} \cite{btth}: &~~& L,R:H:T ~=~ (0.457\pm0.010):(0.522^{+0.038}_{-0.043}):(0.402^{+0.037}_{-0.034});
\label{vab1}\\
\mathrm{Ref.} \cite{hug}: &~~& L,R:H:T ~=~ (0.441\pm0.006):(0.508^{+0.037}_{-0.039}):(0.388^{+0.033}_{-0.030});
\\
\mathrm{Ref.} \cite{fb}: &~~& L,R:H:T ~=~ (0.515\pm0.005):(0.573^{+0.030}_{-0.036}):(0.453^{+0.036}_{-0.033}).
\label{vab3}
\end{eqnarray}
Comparing them with the experimental value found by Belle collaboration\cite{bel3}, $F_L^{D^*}$ =$0.60\pm0.08\pm0.04$, we conclude that the evaluation according to the $T$-interaction differs from data by $\sim ~2\sigma$, while the one of the $H$-interaction is a bit favored over those of the $L$- and $R$-interactions, which are compatible with experiment within $\sim ~1\sigma$.

Moreover, owing to angular momentum conservation, this observable equals $ F_L^{{\cal B}^*}$ in the $B \to D^*$ decay; therefore, it is interesting to compare the values (\ref{vab1}) to (\ref{vab3}) with (\ref{algnm}), which concern the $\Lambda_b \to \Lambda_c$ decay. We note that these last results are not so different than (\ref{vab3}), which are somewhat related to experimental data.

\section{Summary and Discussion}

Here we summarize the main results of our paper, then we shortly illustrate alternative approaches and suggest new measurements and calculations.

1) We assume that the features of the semi-leptonic decays are practically independent of the specific structure of the hadrons involved. This implies that the ratio ${\cal R}/{\cal R}^{SM}$ is approximately equal for the different semi-leptonic decays, 
giving rise to results that agree with other, more sophisticated, calculations\cite{fb}. It has also a remarkable predictive power for some decays, according  to Eqs. (\ref{prdn}), and it helps explaining quite simply some features of preceding phenomenological analyses of $B$ decays. 

2) As a further consequence of our assumption, our calculations on the $\Lambda_b\to \Lambda_c$ decay have to be combined with previous analyses of the $B\to D$ and $B\to D^*$ decays, which induces serious constraints on the parameters that characterize each possible NP interaction. Besides, if we regard such an interaction as a perturbation of the SM\cite{ig2} - as it appears from data and from Eqs. (\ref{asspt}) and (\ref{newdlt}) - we are led to indicating the $L$-interaction as responsible for the LFU violation, since the other Dirac operators contradict this standard. In particular, the $T$- and the $(S+P)$-interaction are excluded also by previous analyses\cite{ddut,adm,tw}. The $H$-interaction appears slightly favored by the data of $F_L^{D^*}$; however, it demands a large coupling, which is compensated by a negative $\cos \psi$. Analogously, the $R$-interaction, whose phase is automatically fixed by the fit to the data, entails a sizable coupling. Our conclusion agrees with that of our preceding article\cite{dsfa} and with other authors\cite{adm, Jan,hug,bd,kks2}. Under the assumption of a single NP operator, it matches up also with the analysis of ref. \cite{ikw}, while ref. \cite{hu1} admits, alternatively, also the $T$-interaction.

3) On the contrary, several authors assume more NP operators, in view, for example, of more flavor issues\cite{ikw,ig2}.

a) Some model independent analyses demand at least two operators\cite{ivn,As2,chdr,celi}, while others find that no combination of all operators may explain presently available data\cite{AsS,trn}.

b) Looking at the dynamics which produces the anomaly, the leptoquark model - which implies two operators - is especially intriguing in explaining the LFV, as the quark-quark interactions are not involved in the tensions with the SM\cite{iiv}; according to the results of the semi-leptonic decays where light leptons are observed, we conclude that such an interaction must depend on the lepton mass. In particular, the vector leptoquark[78,118-124] (see also refs. 201 to 209 of \cite{ikw}) demands a prevalence of the $L$-operator: a best fit to data yields a $22\%$ of the $H$-interaction\cite{ikw}.
Other authors find that data may be explained both by the vector and scalar leptoquark model\cite{che,frts}. Two scalar leptoquarks are assumed by refs. \cite{darl,yyy} and \cite{cvn,tw}, but while the first two conclude that the $L$-operator dominates over $S-P$, the last two propend for a prevalence of the $T$-operator. Refs. [127-129]
explain the anomaly by a single scalar leptoquark, but they are contradicted by Bansal {\it et al.}\cite{bck}. Last, ref. \cite{alda} remarks that the vector leptoquark model does not explain the anomalies in the muon sector.
 
4) In order to discriminate between our conclusion and those of other authors, we propose various tests. For instance, the angular distribution of the decay 
$\Lambda_b \to \Lambda_c \tau {\bar \nu}_{\tau}$ offers, in principle, the richest font of information\cite{rdt}, especially - but not exclusively - with polarized $\Lambda_b$: indeed, it allows to determine five observables, which assume different values according to the Dirac operator that describes NP (see Table 2); however, as shown by Eq. (\ref{tvas}), the size of the TRV asymmetry - predicted, according to our calculations, for all NP interactions but $L$ - depends crucially on the polarization of $\Lambda_b$. On the other hand, the parameter $\beta_L^{\Lambda_c}$, which may be determined independently of the $\Lambda_b$ polarization, is rather sizable and sensitive to the type of NP interaction. Last, we stress once more that the tests which we propose for this decay can be performed even without knowing the momentum distribution of the $\tau$ lepton, a feature that is shared also by the decay $B\to D^* \tau {\bar \nu}_{\tau}$. 

5) Both decays $B\to D^*$ and $\Lambda_b \to \Lambda_c$ are sensitive to the longitudinal polarization of the $\tau-{\bar \nu}_{\tau}$ system. Our theoretical predictions of this observable are different according to the kind of NP interaction\cite{fdl24} and also according to the decay that we consider. However, we have found an indication of consistency between the predictions based on the two decays, which seems to agree with the consequences of our assumptions. 
 
6) The possible deviations from the SM of decays which involve light leptons can be tested by means of the fully differential distributions. This opens to the possibility of further tests for discriminating among the different NP interactions and for detecting possible TRV. Indeed, a NP interaction which differs from $L$ may involve this violation. In particular, TRV asymmetries can be explored through azimuthal distributions in the decays $B\to D^* \ell {\bar\nu}_\ell$ and $\Lambda_b \to \Lambda_c \ell {\bar\nu}_\ell$\cite{gjl}. 

Last, we give some suggestions for future, in-depth, analyses.  

- The measurements of the ratios ${\cal R}$ for the decays $\Lambda_b \to \Lambda_c \tau {\bar \nu}_{\tau}$ and $B_c\to J/\psi \tau {\bar \nu}_\tau$ should be improved, in order to confirm the size of the anomaly found for the $B\to D^{(*)} \tau {\bar \nu}_{\tau}$ decay.

- Precise measurements of the decays $B_c\to l {\bar \nu}_l$ and $B\to X_c l {\bar \nu}_l$ would help in delving into NP.

- A wealth of polarized $\Lambda_b$ would be suitable as well, which could be obtained either through $e^+e^-$ $\to$ $Z^*$ $\to$ $b\bar{b}$[132-134], or in $p-p$ collisions\cite{cms}; in the latter case, it is befitting to look for $\Lambda_b$ resonances produced near the forward direction\cite{ko,ak}.

- Improvements in determining the form factors for the different semi-leptonic decays are very important, since they will reduce, thanks to our assumption or to sum rules\cite{fb,psr}, the ambiguities on the NP interaction that we are looking for.


\vskip 0.50cm



\setcounter{equation}{0}
 \renewcommand\theequation{A. \arabic{equation}}

 \appendix{\large \bf Appendix A}
\vskip 0.30cm

The kinematic variables, which in Sect. 6 were calculated in given reference frames (RF), are converted into Lorentz invariant expressions. The notations are the same as in the text.
\vskip 0.30cm
\centerline{\bf Leptonic Variables}
\vskip 0.30cm

These variables are expressed in the $l-{\bar{\nu}_l}$ RF. Since we assume $m_{\bar{\nu}_l}$ = 0, we get  
\begin{equation}
Q ~ = ~ E_l+p_l,  ~~~~~~ E_l^2 ~ = ~ p_l^2+m_l^2 ~~~ \implies ~~~ p_l ~ = ~ \frac{1}{2Q}(Q^2-m_l^2).
\end{equation} 
Moreover,
\begin{eqnarray}
F_l &=& \sqrt{p_l(E_l+m_l)} = \frac{1}{2}(Q+m_l)\sqrt{v}, ~~~~~~ v = 1-\frac{m_l^2}{Q^2}, 
\\
\chi_l &=& \frac{p_l}{E_l +m_l} = \frac{Q^2-m_l^2}{(Q+m_l)^2}, \ ~~~~~~ \ \ ~~~~~~ \ \ ~~~~~~ \
\end{eqnarray} 
whence
\begin{equation}
F_l(1-\chi_l) = m_l\sqrt{v}, ~~~~~~ F_l(1+\chi_l) = Q\sqrt{v}.
\end{equation} 
\vskip 0.30cm
\centerline{\bf Hadronic Variables}
\vskip 0.30cm

Such variables were calculated in the $\Lambda_b$ RF. We have
\begin{equation}
m_{\Lambda_b} =  p_0+E_{\Lambda_c},  ~~~ p_0^2 = Q^2+p^2 ~~~ E_{\Lambda_c}^2 =  p^2+ m_{\Lambda_c}^2.
\end{equation}
Then, we deduce 
\begin{equation}
p_0 =\frac{1}{2 m_{\Lambda_b}}(M_+M_- +Q^2), ~~~ p =\frac{\sqrt{Q_+Q_-}}{2 m_{\Lambda_b}}  ~~~ E_{\Lambda_c} =
=\frac{1}{4 m_{\Lambda_b}}(Q_+ +Q_-),
\end{equation} 
where
\begin{equation}
M_{\pm}=  m_{\Lambda_b}\pm m_{\Lambda_c}, ~~~~~~ Q_{\pm}= M_{\pm}^2-Q^2.
\end{equation}
Similarly, 
\begin{equation}
E_{\Lambda_c}\pm m_{\Lambda_c} =  \frac{Q_{\pm}}{2 m_{\Lambda_b}}.
\end{equation}
Then
\begin{equation}
F = \sqrt{2m_{\Lambda_b}(E_{\Lambda_c}+m_{\Lambda_c})} =  \sqrt{Q_+},
\end{equation}
\begin{equation}
\chi = \frac{p}{E_{\Lambda_c}+m_{\Lambda_c}} = \sqrt{\frac{E_{\Lambda_c}-m_{\Lambda_c}} {E_{\Lambda_c}+m_{\Lambda_c}}} = \sqrt{\frac{Q_-}{Q_+}}
\end{equation}
\vskip 1cm
and  
\begin{eqnarray}
F\chi &=& \sqrt{Q_-}, ~~~ F(1\pm\chi) = \sqrt{Q_+}\pm\sqrt{Q_-},  
\\
F(p+\chi p_0) &=& M_+\sqrt{Q_-}, ~~~ F(p_0+\chi p) = M_-\sqrt{Q_+}.
\end{eqnarray}

\vskip 0.30cm
\setcounter{equation}{0}
 \renewcommand\theequation{B. \arabic{equation}}
\appendix{\large \bf Appendix B}
\vskip 0.30cm

Here we list the various parametrizations of the form factors ${\cal F}_i$ and ${\cal G}_i$, $i$ = 1,2,3, for the $\Lambda_b$ $\to$ $\Lambda_c$ transition, to be inserted into Eq. (\ref{pscl}) for our calculations. 

1) Isgur-Wise (IW\cite{iw}) form factors[104-106] (see also \cite{bliw}):
\begin{equation}
{\cal F}_2 = {\cal F}_3 = {\cal G}_2 = {\cal G}_3 = 0, ~~~~
{\cal F}_1 = {\cal G}_1 = \zeta, \label{iw0}
\end{equation}
In particular, we assume two different parametrizations,
\begin{eqnarray}
\zeta_1 &=& 1-\rho^2[\omega-1]+1/2\sigma^2[\omega-1]^2, \cite{klw,blr}
\\
\zeta_2 &=& [2/(\omega+1)]^{\alpha}, ~~~ \alpha=3.5+1.2/\omega\cite{rhk}.
\end{eqnarray}

2) Single pole approximation form factors:

a) with sum rules\cite{dec}:
\begin{equation}
{\cal F}_i = {\cal G}_i = \frac{{\cal F}_i(0)}{(1-Q^2/M_i^2)}, ~~ i=1,2, ~~ {\cal F}_3 = {\cal G}_3 = 0,
\label{monop}
\end{equation}
where $M_i$ are given mass poles;

b) extrapolated from lattice calculation\cite{de}:
\begin{equation}
{\cal F}_i = \frac{b_i+c_i z}{1-Q^2/M_i^2}, ~~ z=\frac{\sqrt{M_+^2-Q^2}-\sqrt{M_+^2-M_-^2}}{\sqrt{M_+^2-Q^2}+
\sqrt{M_+^2-M_-^2}},
\label{pole}
\end{equation}
where $M_{\pm}$ = $M_{\Lambda_b}\pm M_{\Lambda_c}$. 

3) Three-parameter form\cite{gu8,ke,lly}: 
\begin{equation}
{\cal F}_i = \frac{{\cal F}_i(0)}{(1-r)(1-b_{1i}r + 
b_{2i}r^2)}, ~~ r=Q^2/M_{\Lambda_b}^2.
\label{dip1}
\end{equation}

4) Quasi-potential approach\cite{az,fg}:
\begin{equation}
{\cal F}_i = \frac{{\cal F}_i(0)}{1-a_ir+b_ir^2+c_ir^3+d_ir^4}.
\label{dip2}
\end{equation}

The ${\cal G}_i$ that correspond to the form factors (\ref{pole})-(\ref{dip2}) are parametrized analogously to the ${\cal F}_i$.

\end{document}